\documentclass[sigconf]{acmart}
\usepackage{booktabs} 
\usepackage{hyperref}
\usepackage{url}
\usepackage{multirow}
\usepackage{cleveref}
\usepackage{float}
\usepackage{balance}
\usepackage{microtype}
\usepackage{soul}
\usepackage{color}
\usepackage{graphics,graphicx}
\usepackage{epstopdf}
\usepackage{colortbl}
\usepackage{diagbox}
\usepackage{tabularx}   
\usepackage[justification=centering]{caption}
\usepackage[inline]{enumitem}
\usepackage{acmart-taps}

\newcounter{posts}
\newcommand{\post}[0]{\noindent\refstepcounter{posts}\theposts. }

\setlist[description]{font=\normalfont\itshape, leftmargin=1em, labelindent=0em}

\usepackage{tikz}

\copyrightyear{2025}
\acmYear{2025}
\setcopyright{cc}
\setcctype{by}
\acmConference[CHIIR '25]{2025 ACM SIGIR Conference on Human Information Interaction and Retrieval}{March 24--28, 2025}{Melbourne, VIC, Australia}
\acmBooktitle{2025 ACM SIGIR Conference on Human Information Interaction and Retrieval (CHIIR '25), March 24--28, 2025, Melbourne, VIC, Australia}
\acmDOI{10.1145/3698204.3716456}
\acmISBN{979-8-4007-1290-6/2025/03}

\begin{document}

\title{Search Changes Consumers' Minds}
\subtitle{How Recognizing Gaps Drives Sustainable Choices}

\author{Frans van der Sluis}
\affiliation{\department{Department of Communication}
  \institution{University of Copenhagen}
  \city{Copenhagen}
  \country{Denmark}
}
\email{f.vandersluis@acm.org}
\author{Leif Azzopardi}
\affiliation{\department{Departement of Computer and Information Sciences}
  \institution{University of Strathclyde}
  \city{Glasgow}
  \country{United Kingdom}
}
\email{leifos@acm.org}

\begin{abstract}
Despite a growing desire among consumers to shop responsibly, translating this intention into behaviour remains challenging.
Previous work has identified that information seeking (or lack thereof) is a contributing factor to this intention-behaviour gap.In this paper, we hypothesize that searching can bridge this gap -- helping consumers to make purchasing decisions that are better aligned with their values.
We conducted a task-based study with 308 participants, asking them to search for information on one of eight ethical aspects regarding a product they were actively shopping for.
Our findings show that actively searching for such information led to an overall increase in the importance participants' assigned to ethical aspects.However, it was the recognition and understanding of ethical considerations, rather than ethical intentions or search activity, that drove shifts towards more responsible purchasing decisions. 
Participants who acknowledged and filled knowledge gaps in their decision making 
showed significant behaviour change, including increased searching and a stronger desire to alter their future shopping habits. We conclude that responsible consumption can be considered a partial information problem, where awareness of one's own knowledge limitations may be the catalyst needed for meaningful consumer behaviour change.
\end{abstract}



\maketitle

\section{Introduction}\label{sec_introduction}

Every day, millions of consumers search online, trying to find products that balance their needs, wants, and budget. 
While price and quality continue to be critical factors, a growing number of shoppers are now considering the ethical and social implications of their purchases~\citep{casais2022,carrigan2004better_shopping}. 
This trend has given rise to the ``\textit{socially responsible consumer}'', who actively seeks out products that reflect their commitment to particular ethical concerns such as environmental impact, sustainability, labour rights, etc.~\citep{azzopardi2024src}. 
Ethical consumerism and the socially responsible consumer has gained significant attention across various industries, from food to fashion, as consumers begin to demand such credentials of their purchases \citep{desio2024,badhwar2024}.

Despite this growing interest, there exists a persistent and well-documented \textit{intention-behaviour gap} -- the discrepancy between consumers’ expressed desire to make socially responsible purchases and their actual behaviour~\citep{casais2022,CarringtonLostGap}. 
Key contributors to this gap include the availability and cost of ethical products, along with challenges posed by information asymmetries and green-washing ~\citep{CarringtonLostGap,Uusitalo2004EthicalFinland,Wiederhold2018EthicalIndustry,Keller1987EffectsEffectiveness,orourke2016}. 
These challenges result in relevant details about ethical practices being either unavailable or obscured by less pertinent information, and products being falsely marketed as ethical, making it harder for consumers to distinguish between genuinely responsible products and those that only appear so~\citep{nguyen2019,santos2024}.

We hypothesize that consumers could potentially counteract the effects of information asymmetries and green-washing by performing searches to learn more about the ethical and social aspects related to the products they are intending to buy.
However, such comprehensive searches are often impractical due to the significant time and cognitive effort required~\citep{Schmidt1996ASearch,Brynjolfsson1999FrictionlessRetailers}. 
They involve non-trivial navigation through a large variety of complex web sites coupled with abstract and abstruse presentation of ethical information, such as complicated supply chains and legislation.
Additionally, they face the task of assessing source reliability while comparing multiple products across numerous aspects simultaneously ~\citep{Uusitalo2004EthicalFinland,hernandez2023,Zander2012InformationFood}.
These information and decision-making barriers deter many from actively seeking ethical information~\citep{azzopardi2024src}. 
As a result, consumers tend to prioritize easier-to-compare factors such as price and quality, which can lead them to overlook ethical considerations during purchases ~\citep{casais2022,CarringtonLostGap}.

While the cost and effort of searching for ethical product aspects are known to deter consumers from even initiating a search~\citep{azzopardi2024src},
the influence of the search process on responsible decision-making is less well understood. 
Few studies have addressed how effectively search promotes ethical considerations \citep{Grebitus2020SustainableChoices}.
Notably, although access to sustainability information significantly affects consumers with existing ethical concerns~\citep{santos2024,orourke2016}, merely increasing the volume of information has proven insufficient~\citep{Zander2012InformationFood}.
These studies, along with the barriers identified, show that the search process itself matters.
It also raises questions about what makes a search effective in shaping consumer choices, and, how seeking ethical information might influence consumers even if they are not ethically minded.

This paper investigates how actively searching for ethical information influences consumer priorities during purchase decisions. It examines this through a task-based study in which 308 participants searched for information on one of eight ethical aspects of products, where we consider the following research questions:

\aptLtoX[graphic=no,type=html]{\begin{enumerate}%
    \item[RQ1] \label{rq1} Whether actively seeking ethical information affects consumers’ purchasing intentions, i.e., does search change consumer minds?
    \item[RQ2] \label{rq2} Whether -- (a) ethical intentions, (b) search activity, or (c) perceived barriers  -- result in the greatest change in the importance of ethical considerations?
\end{enumerate}
}{\begin{enumerate}[label=RQ\arabic*]
    \item \label{rq1} Whether actively seeking ethical information affects consumers’ purchasing intentions, i.e., does search change consumer minds?
    \item \label{rq2} Whether -- (a) ethical intentions, (b) search activity, or (c) perceived barriers  -- result in the greatest change in the importance of ethical considerations?
\end{enumerate}}
We hypothesize that explicitly searching for and finding ethical information related to their product could 
narrow the intention-behaviour gap, leading to better-aligned purchasing decisions. 
Our study employs a correlational design to explore how the search process impacts decision-making priorities, supplemented by a qualitative analysis of participants' responses for a richer understanding.%

 \section{Background} \label{sec_background}
To position this work, we will provide an overview of how search is central to consumer purchasing decisions, followed by the rise of ethical consumers before detailing the intention-behaviour gap -- where we hypothesize that searching could help allievate this gap.

\subsection{Consumer Purchasing Models}\label{bg:purchase_models}
Most purchasing decisions involve searching for information about available products, followed by evaluating this information to make a purchase decision. This has lead to the development of a number of similar models ascribing consumer purchasing behaviour~\citep{Bloch1986ConsumerFramework,Engel1990CustomerBehavior,RowleyProductPropositions,Punj1983AnMaking,Ke2016SearchProducts,Schmidt1996ASearch}. 
These models generally describes a series of stages, including need recognition, information search, evaluation of alternatives, purchasing decision, and post-purchase evaluation~\citep{Engel1990CustomerBehavior}. This structured approach emphasizes that consumers typically start by recognizing their needs and then actively seek information about products that meet those needs. Such behaviors are not merely driven by the product’s attributes but can be significantly influenced by ethical considerations~\citep{Ajzen1985FromBehavior}.

\begin{figure}
\includegraphics[width=4.4cm]{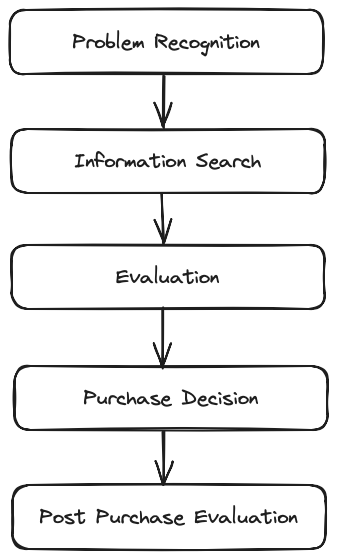}
\caption{A simple model of the consumer buying process by \citet{Engel1990CustomerBehavior}.}\label{fig:buying_process}
\end{figure}

Despite the availability of information, consumers often face challenges during the search and evaluation phases that can impede their decision-making effectiveness. Specifically, they may rely on incomplete or imperfect information due to the following factors:
\begin{enumerate}
    \item The complexity and volume of information available across various platforms can lead to cognitive overload, making navigation through these resources daunting~\citep{Brynjolfsson1999FrictionlessRetailers}.
    \item Information asymmetries exist, where sellers may not disclose crucial details about their products, complicating the consumer's ability to make informed choices~\citep{Akerlof1970TheMechanism,Jacoby1974BrandLoad}.
    \item With numerous alternatives, consumers frequently experience decision fatigue, as the sheer variety of options can overwhelm them~\citep{Keller1987EffectsEffectiveness}.
\end{enumerate}

Consequently, rather than enhancing decision-making, the abundance of information can hinder it~\citep{Keller1987EffectsEffectiveness,Schleenbecker2015InformationCoffee}. Although branding and labeling initiatives (e.g., Fair Trade, Energy Efficiency) aim to reduce information asymmetry and assist consumers in their decision-making processes~\citep{WardInternetBranding,Young2010SustainableProducts,Annunziata2011ConsumersProducts}, 
consumers often adopt selective search strategies that simplify their searches~\cite{Zander2012InformationFood}. 
Given that searching, evaluating and comparing products and the associated information can be quite costly (in terms of time and effort), consumers frequently make decisions based on incomplete and imperfect information~\citep{WardInternetBranding,Ke2016SearchProducts}. 
This often leads to omission neglect, where consumers fail to consider certain aspects because the information is not immediately available or easy to find~\citep{Kardes2006DebiasingNeglect,Pfeiffer2014EffectsEvaluation,sanbonmatsu2003}.

\subsection{Socially Responsible Consumers}\label{bg:src}
In today’s marketplace, consumers are increasingly aware of the social and environmental implications of their purchasing choices. This heightened awareness has led to a growing trend toward ethical consumerism, where individuals prioritize the broader impact of their purchases alongside personal preferences~\citep{Ellen1991TheBehaviors,CarringtonLostGap,Jones2022TheDifference,Davies2016ConsumerConsumption,Sheth2011MindfulSustainability}.

A ``\textit{socially responsible consumer}'' is an individual who makes purchasing decisions with a focus on the broader social and environmental impact of their choices, in addition to their personal needs and preferences~\cite{Ellen1991TheBehaviors,CarringtonLostGap,Jones2022TheDifference,Davies2016ConsumerConsumption, Sheth2011MindfulSustainability}\footnote{Socially responsible consumers are also referred to as a mindful consumers or ethical consumers.}. 
According to Carrigan et al.~\cite{carrigan2004better_shopping}, \textit{responsible consumerism} broadly stands for ``\textit{the conscious and deliberate choice to make certain consumption choices due to personal and moral beliefs}'' and thus responsible consumers prioritize products and services that align with their personal values and beliefs~\cite{CarringtonLostGap,Jones2022TheDifference}. 
These may include beliefs and values about the environment, human and animal rights, community involvement, social justice, governance, etc.~\cite{Jones2022TheDifference,Shaw2006FashionChoice}.

Social responsibility, however, is perceived differently by each consumer, meaning that an aspect important to one person might not be important to another. Hasanzade et al.~\cite{Hasanzade2018SelectingShopping} surveyed 249 consumers in Germany and found that most participants were ethically minded consumers (54\%), while the other participants were concerned with price (12\%) and price-quality (34\%). Of those ethically minded, most were concerned about animal rights, followed by labour/human rights, and then environment protection. Casais et al.~\cite{Casais2022ThePriorities} surveyed 364 consumers in Portugal, of which most considered themselves as socially responsible consumers, reported that they were concerns about labour/human rights (31\% ), environmental issues (23\%), animal rights/welfare (17\%), and all three (27\%). The variety of ethical concerns translates into highly varied purchasing decisions.  For example, a consumer who consciously avoids products made in regions associated with slave/forced labour may be indifferent to the ecological or sustainability implications of their choice. Therefore, not all consumers will attribute the same importance to the different issues/aspects associated with being ethical and socially responsible.

Nonetheless, socially responsible consumers aim to use their purchasing power as a means to promote positive change. 
They believe that by supporting companies and products that align with their values and sustainability goals, they can influence businesses to be more socially and environmentally responsible~\cite{Liu2018TheIntention}. 
This consumer mindset has grown in prominence with increasing awareness of global environmental issues, labor rights, and corporate responsibility, leading to the growth of various certifications and labels to help consumers identify products that meet these criteria~\cite{Jones2022TheDifference}. 
Socially responsible consumers tend to establish an identity rooted in their ethical purchasing decisions, occasionally making personal sacrifices~\cite{Papaoikonomou2018LookingApproach}. 
And, they often tend to communicate their role as advocates for a more sustainable society to others~\cite{Casais2022ThePriorities}.

\subsection{Intention-Behaviour Gap}\label{bg:gap}
The \textit{Theory of Planned Behaviour} seeks to provide an explanation of behaviour by considering how the individual's attitudes and norms influences their intentions and subsequently their behaviour in a casual sequence~\cite{Ajzen1985FromBehavior}.
People's intentions to purchase socially responsible products, however, do not always align with their behaviors. For example, of 81 self-declared green consumers, 30\% reported that they were very concerned about environmental issues but they struggled to translate this into their purchasing decisions~\cite{Young2010SustainableProducts}. 
This has been referred to as the \textit{Intention-Behavior Gap}~\cite{Ajzen1985FromBehavior}. Within the literature on consumer purchasing, this misalignment has been of great interest, especially regarding ethical concerns~\cite{Moon2004ConsumerHypotheses,Ozcaglar-Toulouse2006InFrance,RowleyProductPropositions}.

Several studies have identified key barriers contributing to the intention-behavior gap. 
Consumers often struggle to access reliable information on ethical aspects -- such as sustainability or labor practices -- since companies may either withhold such data or obscure it within marketing content~\citep{Uusitalo2004EthicalFinland}. 
Ethical products are also perceived as more expensive or less available, deterring price-sensitive shoppers~\citep{Wiederhold2018EthicalIndustry}. 
Moreover, the overwhelming amount of product information online can cause consumers to focus on simpler factors like price and quality, even if they have ethical intentions~\citep{Keller1987EffectsEffectiveness}. 
Carrington et al.~\citep{CarringtonLostGap} emphasize that this gap is also influenced by habitual purchasing behaviors and alternative personal values, with ethical concerns often taking a back seat to convenience and cost. Consequently, despite growing awareness of ethical issues, many consumers continue to prioritize ease and affordability over responsible consumption in their decision-making. 

When consumers search for ethical information, they face additional barriers. The sheer volume of available data often leads to cognitive overload, making it harder to navigate ethical considerations~\citep{Brynjolfsson1999FrictionlessRetailers}. Compounding this issue is omission neglect, where consumers unintentionally overlook ethical aspects because they are not prominent or easy to find~\citep{hernandez2023}.
To reduce the complexity of the decision-making process, many consumers adopt simplifying strategies, which can lead them to disregard ethical information entirely~\citep{Zander2012InformationFood}. 
In line with this, in a recent survey of 302 searchers, Azzopardi and van der Sluis~\cite{azzopardi2024src} found that participants frequently prioritized primary factors such as price and quality over ethical considerations, highlighting a growing gap between consumer intentions and actual search behavior. 
Searchers reported challenges including a lack of awareness about ethical issues, difficulty accessing relevant information, trouble interpreting key details, distrust in the information available, and the complexity of comparing options. Even when participants engaged with ethical information, they encountered these difficulties throughout the different stages of purchasing.

Despite these barriers, there is evidence that information-seeking can help bridge the intention-behavior gap. 
When confronted with ethical information during the search process, consumers often become more concerned and perform more extensive searching~\cite{Zander2012InformationFood}.
Consumers who actively search for sustainability information tend to have higher purchase intentions for ethically aligned products~\citep{orourke2016}. 
However, this effect is most pronounced among those who are already ethically engaged, leaving open the question of how effective search is for less motivated consumers.
Moreover, while more accessible information could help close this gap, merely increasing the volume of information isn’t enough. It must be presented in a way that seamlessly integrates into the decision-making process~\citep{Zander2012InformationFood}. 
These findings suggest that search may mitigate or overcome some of the identified barriers,
but the specific mechanisms by which search can influence responsible consumption are not yet fully understood.

 \section{Method}\label{sec_method}
The research employed a mixed method design using a cross-sectional survey to investigate the impact that searching for ethical information has on responsible consumption choices. Below we explain our study design in detail, the demographics of our participants along with how we analysed the data collected.

\subsection{Design}\label{met:design}
The study was composed of three parts. Pre-task questionnaire, followed by a search task, and then a post-task questionnaire. 
The study design and questionnaires were adapted from a previous survey~\citep{azzopardi2024src}.
Following our research questions, the main dependent measure was \textit{importance change}, capturing shifts in participants' ethical valuations resulting from the search process. A pre-test/post-test setup, despite potential demand characteristics or social desirability bias, was favoured to distinguish participants' initial intentions from the effects of the search task, enabling us to track changes in importance. This design also provided real-time insight into how new information encountered during the search could influence decision-making priorities, in line with the study’s focus on the role of information in responsible consumption.

The study was conducted in compliance with our organisation's ethical standards and we received approval from our Ethics Committee ({Application No. 2680}).
Participants' informed consent was obtained, and their privacy and confidentiality were maintained throughout the research process. Demographics data was collected via the Prolific.
The survey was administered online using the Qualtrics survey platform while the search task was completed via an instrumented web search engine (powered by Google). 
All search queries and page opens were tracked directly within the search interface, as detailed in \citep{vandersluis2025ipm}.
Participants were informed about the research study and its objectives through an information / consent form. 
They were explicitly told about the voluntary nature of participation, and their ability to withdraw from the study at any time without consequences. The survey's estimated completion time was 20 minutes. Participants were compensated in line with national wage guidelines.

\subsubsection*{Part 1: Purchase Questions}
Part 1 asked participants to describe their current purchasing decision.
By focusing on tangible, real-world purchase rather than a hypothetical scenario, the study aimed to mitigate potential social desirability bias. 
This approach also meant that the search task was aligned with their actual task of shopping -- and therefore could impact upon their decision.
For the full list of questions, please refer to the appendix (Questions \hyperref[q:confirmation]{Q1} -- \hyperref[q:alternatives]{Q7}, Supplement \hyperref[app:questions]{B}).

Participants first provided a description of the product they were considering purchasing and selected its category. These categories were sourced from Statista.com\footnote{\url{https://www.statista.com/statistics/276846/reach-of-top-online-retail-categories-worldwide/}, last accessed on February 24, 2025}. 
Participants then provided information about the stage of the purchasing decision, the duration of the search process, and how many alternatives they had considered to date.
These questions aimed to assess the significance of their reported purchase.
In the creation of these questions, care was taken to adhere to survey guidelines, ensuring that the questions requested concrete numerical responses, avoided ambiguity, and provided a balanced set of answer options \citep[][ch. 11]{clark2021bryman}.

\subsubsection*{Part 2: Search Task}

The second part of the study asked participants to spend approximately 10 minutes searching for a given ethical aspect associated with the product they were interested in purchasing. One of eight ethical aspects -- defined as the combined ethical, social, and environmental dimensions -- was randomly assigned.
For each aspect, a brief explanation was shown along with few example questions (see Table~\hyperref[app:aspects_overview]{S1}, Supplement \hyperref[app:aspects]{A}).

Before the search task, participants were asked to evaluate the importance of the assigned ethical aspect in their decision-making (Question \hyperref[q:importance]{Q8}, Supplement \hyperref[app:questions]{B}). This served as a pre-test measure to capture their initial perceptions of the aspect's relevance.
Participants were also asked whether they had considered or specifically searched for the ethical aspect during previous searches (Questions \hyperref[q:considered]{Q9} and \hyperref[q:searched]{Q10}), to gauge their prior engagement with the issue.

Following the search task was a custom-developed 16-item questionnaire (Questions \hyperref[q:keydecision]{Q14} -- \hyperref[q:reconsidering]{Q30}, Supplement \hyperref[app:questions]{B}) about consumers search experience.
The design of this questionnaire is grounded in previous studies on responsible consumer information-seeking challenges \citep{azzopardi2024src}
and derived from consumer purchase process models~\citep{Bloch1986ConsumerFramework,Engel1990CustomerBehavior,RowleyProductPropositions,Punj1983AnMaking,Ke2016SearchProducts,Schmidt1996ASearch}. 
We used both negatively and positively phrased questions to enhance the reliability of the responses.
By structuring the survey around the phases of the consumer purchase process, we aimed to capture the barriers that responsible consumers face during online product searches and their decision-making processes:
\begin{description}
    \item[Problem Recognition]: Questions \hyperref[q:keydecision]{Q14} -- \hyperref[q:moreresearchneeded]{Q17} assessed whether participants recognized the relevance of ethical aspects and identified knowledge gaps in their purchase decisions.
    
    \item[Information Search]: Questions \hyperref[q:easyfind]{Q18} -- \hyperref[q:timeconsuming]{Q21} explored the availability and accessibility of ethical information, focusing on any difficulties participants encountered while searching.
    
    \item[Information Evaluation]: Questions \hyperref[q:sensemaking]{Q22} -- \hyperref[q:toomuchinfo]{Q25} examined participants' ability to understand and trust the information, revealing issues with interpretation and reliability.
    
    \item[Decision-Making]: Questions \hyperref[q:wontuse]{Q26} -- \hyperref[q:reconsidering]{Q30} addressed challenges like information overload and making trade-offs between ethical aspects and other factors like price or quality.
\end{description}

\subsubsection*{Part 3: Search Impact and EMCB Scale}
Part 3 assessed the impact of the search task on participants' ethical considerations. Participants first reassessed the importance of the assigned ethical aspect after search (Question \hyperref[q:updatedimportance]{Q31}, Supplement \hyperref[app:questions3]{B}).
They were then asked open-ended questions regarding what they learned from the search, how it influenced their decision-making, and what challenges they faced when searching (Questions \hyperref[q:learning]{Q32} -- \hyperref[q:challenges]{Q34}).

Following these questions, participants completed the standardized {\textit{Ethically Minded Consumer Behavior} (EMCB)} scale~\citep{SUDBURYRILEY20162697}, providing a quantitative assessment of their eco-friendly consumption practices. 
This scale was placed at the end of the survey to avoid influencing participants' responses to other questionnaire items.
The EMCB scale is extensively tested among consumers in multiple countries, demonstrating its cross-cultural validity~\citep{SUDBURYRILEY20162697}.

\subsection{Participants}\label{met:participants}
The study involved 320 participants who were recruited through the online platform Prolific. Eligible participants were required to be 18 years or older, capable of giving informed consent, and asked to confirm that they were in the process of making a purchase valued at approx. \$100/€100 or more.
Of the 320 individuals who entered the survey, 9 didn't fully complete the survey, 2 who didn't given consent, and one stated they didn't qualify, resulting in 308 participants.
Of the $308$ participants who fully completed the study, $40\%$ were female, $60\%$ were male. Their ages ranged from $18$ to $70$, with the average age being 31.7 (+/-9.90) years old.

Participants were of diverse backgrounds and geographic locations, enhancing the study's generalizability. The majority of participants were from the United Kingdom (88), followed by the United States (46), South Africa (41), Poland (32), and Canada (15). In terms of employment status, most participants were employed full-time (158), with smaller groups working part-time (47), actively seeking employment (39), with the remainder indicating other employment statuses.

Participants reported purchasing a range of products, with the most common categories being Consumer Electronics ($55.52\%$), Clothing, Shoes, Fashion Accessories ($14.94\%$), Household Appliances \& Goods, Furniture ($14.61\%$), Sports, Recreation, Hobbies ($5.52\%$), and Other ($3.57\%$).
Participants were at various stages of their shopping task: $19.81\%$ were just starting to look for products that met their needs, $9.74\%$ were searching for different alternatives, $22.08\%$ were comparing and researching specific alternatives, $12.34\%$ were deciding which alternative to purchase, and $34.42\%$ were looking for the best possible deal for the product they selected. 
The remaining $1.30\%$ stated that they already purchased the product.

In line with their shopping stage, $8.12\%$ had searched for less than an hour, $33.12\%$ had searched for 1-2 hours, $32.79\%$ had searched for 2-4 hours, $15.26\%$ had searched for 5-8 hours, and $10.06\%$ had searched for more than 9 hours.
The duration of their search extended over the course of a day ($11.04\%$), a week ($55.19\%$), a month ($22.08\%$), and $11.36\%$ searched for longer than a month.
Most participants considered 2-3 alternatives ($62.66\%$), $24.03\%$ considered 4-5 alternatives, and $10.06\%$ considered only one alternative. A small number of participants considered 6 or more alternatives ($2.92\%$).

Taken together, most participants were currently interested in purchasing consumer electronics, had already compared 2-3 alternatives, and had spent 2-4 hours searching over the course of a week for their purchase.

\subsection{Data Analysis and Reporting}\label{met:analysis}
Before proceeding with the analysis, the collected data underwent several pre-processing steps to ensure data quality. These steps included the removal of participants who did not meet the specified criteria and ensuring that no personally identifiable information was associated with the responses. Additionally, 
reverse-coded items were re-scaled, and items within the same factor or scale were averaged to create composite scores.

Data collected from the survey underwent a series of analyses aimed at addressing the research questions. Descriptive statistics, including frequencies and means, provided an overview of the data. Inferential statistical techniques using R were employed, including linear regression, correlational analysis, analysis of variance (ANOVA), and the computation of confidence intervals, to test for relations. 
More specialized analyses included two applications of KMeans clustering: (1) combined with Principal Component Analysis (PCA) to segment and visualize participants based on their EMCB scores (ethical intentions), and (2) combined with Exploratory Factor Analysis (EFA) to uncover latent constructs and group variables in the 16-item questionnaire on search appraisals (Questions \hyperref[q:keydecision]{Q14} -- \hyperref[q:reconsidering]{Q30}, Supplement \hyperref[app:questions]{B}).
Conditional effects analysis using R's emmeans package\footnote{\scriptsize{\url{https://cran.r-project.org/web/packages/emmeans/index.html}}} was employed to examine how ethical intentions (EMCB scores) moderated the effects of search behaviour on importance change.

Open-ended responses from the survey underwent qualitative analysis.
The focus was on identifying illustrative examples of sense-making and gap recognition, both of which were linked to changes in importance valuations. Using the OpenAI's API to assist in the clustering and extraction of comments using GPT4o (See Supplement \hyperref[app:chatgpt]{C} for prompts), we extracted prototypical examples where participants either successfully integrated ethical information into their decisions or struggled to make sense of it\footnote{\scriptsize{Note, that in accordance with our university's policies on the use of generative AI, we used a paid subscription to OpenAI to ensure that participant comments were kept confidential.}}.
AI-assisted codes and AI-extracted examples were reviewed and validated by a human assessor (a co-author) to ensure they were accurate and correct. 
The following section presents the results of these data analyses.

\begin{table}[tb]
    \centering
    \begin{tabular}{lrrrrr}
\toprule 
   & & \multicolumn{2}{c}{\textbf{Prior}} & \multicolumn{2}{c}{\textbf{Change}} \\ 
 \cmidrule(l){3-4} \cmidrule(l){5-6} 
  \textbf{Aspect} & \textbf{Cnt}. & \textbf{Avg}. & \textbf{Std}. & \textbf{Avg}. & \textbf{Std}. \\ 
   \midrule
\hline
  \textbf{Governance} &  44 & 2.98 & 1.24 & -0.12 & 1.09 \\ 
  \textbf{Labour} &  34 & 2.91 & 1.24 & 0.12 & 0.94 \\ 
  \textbf{Sourcing} &  32 & 2.88 & 1.13 & 0.53 & 0.86 \\ 
  \textbf{EcoFriendly} &  39 & 2.67 & 1.20 & 0.42 & 0.76 \\ 
  \textbf{Origin} &  47 & 2.47 & 1.10 & 0.41 & 1.32 \\ 
  \textbf{DEI} &  42 & 2.43 & 1.48 & 0.26 & 1.00 \\ 
  \textbf{SocialImpact} &  43 & 2.37 & 1.29 & 0.45 & 1.11 \\  
  \textbf{Ideology} &  59 & 2.15 & 1.30 & 0.36 & 0.95 \\ 
     \bottomrule
\end{tabular}
 \caption{Counts (cnt.), averages (avg.), and standard deviations (std.) of prior importance and importance change ratings across different aspects.
{\normalfont A one-way ANOVA revealed an overall effect of \emph{Aspect} on prior \emph{Importance}, $F(7, 335) = 2.43$, $p = 0.019$, $\eta^2 = 0.05$.
The effect of \emph{Aspect} on \emph{Importance change} was not statistically significant, $F(7, 315) = 1.45$, $p = 0.186$, $\eta^2 = 0.03$.
Post-hoc pairwise comparisons using Tukey's HSD test showed no significant differences for \emph{Importance change} ($.217 < p \leq 1.00$). For \emph{Importance}, comparisons were non-significant ($.100 < p < 1.00$), with one exception: \emph{Ideology} -- \emph{Governance} with $p = .033$.
        }
    }
    \label{tab:aspects}
\end{table}

\begin{table*}[tb]
    \centering
    \begin{tabular}{lrrrrr}
\toprule 
   & & \multicolumn{2}{c}{\textbf{Prior}} & \multicolumn{2}{c}{\textbf{Change}} \\ 
 \cmidrule(l){3-4} \cmidrule(l){5-6} 
  \textbf{Product Category} & \textbf{Cnt.} & \textbf{Avg.} & \textbf{Std.} & \textbf{Avg.} & \textbf{Std.} \\ 
 \midrule
  \hline
    \textbf{Consumer electronics} & 190 & 2.45 & 1.27 & 0.25 & 1.01 \\ 
    \textbf{Household appliances \& good, furniture, etc. }&  52 & 2.81 & 1.30 & 0.45 & 1.04 \\ 
    \textbf{Clothing, Shoes, Fashion Accessories} &  51 & 3.00 & 1.28 & 0.21 & 1.15 \\ 
    \textbf{Sports, Recreation, Hobbies, etc.} &  22 & 2.14 & 1.08 & 0.60 & 0.99 \\ 
    \textbf{Books, movies, games, toys} &  12 & 2.00 & 0.85 & 0.50 & 0.67 \\ 
    \textbf{DIY \& Garden} &   5 & 2.20 & 1.10 & 0.80 & 1.30 \\ 
    \textbf{Other} &  12 & 3.08 & 1.56 & 0.08 & 1.38 \\   
    \bottomrule
\end{tabular}
 \caption{Counts (cnt.), averages (avg.), and standard deviations (std.) of importance and importance change ratings across different product categories.
{\normalfont A one-way ANOVA revealed an overall significant influence of \emph{Product Category} on \emph{Importance (prior)}, $F(6, 336) = 2.80$, $p = .011$, $\eta^2 = 0.05$.
The effect of \emph{Product Category} on \emph{Importance change} was not statistically significant, $F(6, 316) = 0.92$, $p = .479$, $\eta^2 = 0.02$.
Post-hoc pairwise comparisons using Tukey's HSD test showed no significant differences for \emph{Importance change} (all $p > .80$) or for \emph{Importance (prior)} ($.093 < p < 1.00$), though a marginal difference was observed between \emph{Clothing, Shoes, Fashion Accessories} and \emph{Consumer electronics} ($p = .093$).
        }
    }
    \label{tab:products}
\end{table*}

\section{Results}

Table~\ref{tab:aspects} and \ref{tab:products} summarize participants' importance ratings for individual aspects and product categories, respectively.
Anova tests revealed an overall influence of both ethical aspects (Table~\ref{tab:aspects}) and product categories (Table~\ref{tab:products}) on importance evaluations prior to searching.
Some product categories (e.g., Clothes, Shoes, Fashion Accessories) appeared more closely associated with sustainability considerations than others (e.g., Consumer electronics).
Likewise, some aspects (e.g. Governance) were rated as more important than others (e.g. Ideology).
However, post-hoc analyses failed to offer conclusive evidence for most specific pairwise comparisons,
revealing only marginal differences in some cases.
These findings are consistent with earlier research \citep{azzopardi2024src}, which revealed a large gap between importance evaluations of primary (e.g., price, quality) versus secondary (ethical, social and environmental) aspects,
but less so within secondary aspects nor between product categories.

Participants ratings of the importance of their assigned ethical aspects was shown to significantly increase after searching compared with their pre-search rating (according to a paired t-test comparing pre-search ($\mathrm{M} = 2.57$, $\mathrm{SD} = 1.28$) and post-search importance ratings ($\mathrm{M} = 2.90$, $\mathrm{SD} = 1.22$)).
The test revealed a significant difference between these ratings, $t(322) = 5.28$, $p < .001$, with a mean change in importance of $0.31$ ($\mathrm{SD} = 1.04$), $95\%~\mathrm{CI} [0.19, 0.42]$.
This confirms \ref{rq1} and aligns with earlier studies (see Section \ref{bg:gap}) in showing 
that actively engaging in search will lead to changes in people's minds about the importance of ethical aspects.

Any changes in importance evaluations, however, could not be attributed to specific product categories or ethical aspects,
as described in Table \ref{tab:products} and Table \ref{tab:aspects}, respectively.
This indicates that other factors, such as individual differences or related to the search process, may play a larger role in driving changes in importance ratings.

\subsection{Individual Intentions and Importance (Change)}

\begin{figure*}[tb]
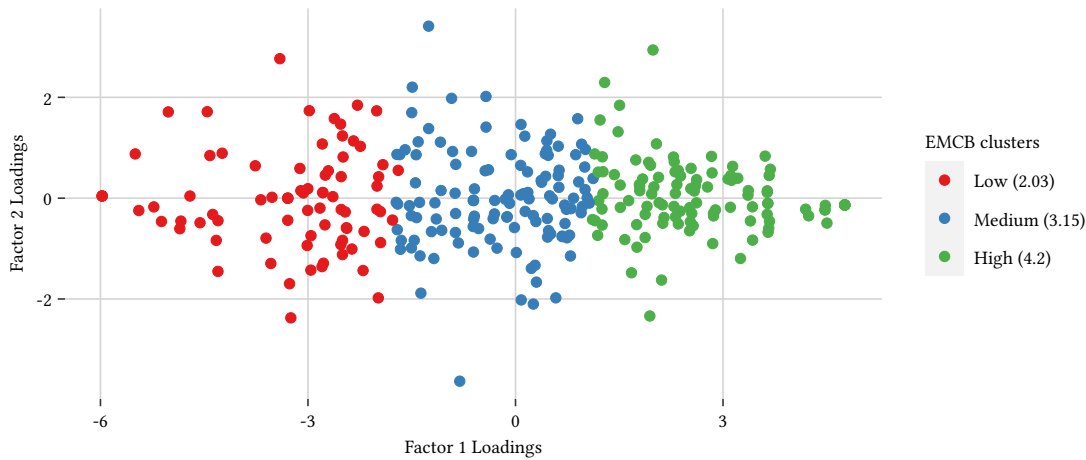

    \centering

     \caption{
         K-Means clustering of participants based on their EMCB answers, visualized using Principal Component Analysis (PCA).
        {\normalfont
         The scatter plot shows three distinct clusters along the first principal component (PCA1), which captures 59.95\% of the variance and highlights the primary differences between groups. The second component (PCA2) explains 7.86\% of the variance, reflecting additional variability among participants. Clustering was performed using standardized scores (mean and standard deviation), with the number of clusters determined by the Elbow Method and silhouette scores (Supplement E). PCA was applied to reduce the data to two dimensions for visualization.
        }
    }
    \label{fig:emcb}
\end{figure*}

Figure \ref{fig:emcb} shows a clustering of participants based on their answers to the EMCB questionnaire.
Of our 308 participants, $22.9\%$ can be identified as least ethically minded or ethically agnostic (low, mean $2.04$), showing little intention or regard for purchasing ethical and socially responsible products. A further $34.9\%$ were moderately ethically minded (medium, mean $3.15$), expressing a mixture of ethical awareness toward certain aspects of social responsibility but not all. For example, they may have expressed concern over environmental issues but not over employment rights, or vice versa. The remaining $31.4\%$ were considered highly ethically motivated consumers (high, mean $4.20$), demonstrating significant concern and behavior toward shopping ethically and responsibly.
This analysis suggests considerable differences between groups of participants in their intentions toward responsible consumption. The outcome aligns with previous research that identified similar distributions of ethically-minded consumers~\cite{Casais2022ThePriorities,Hasanzade2018SelectingShopping}.

The EMCB helps assess the extent to which individual differences in ethical intentions influence importance valuations and their changes.
As shown in Figure \ref{fig:boxplots}a with an ANOVA test, EMCB clusters had a significant and substantial effect on importance ratings, with post-hoc pairwise comparisons confirming differences between clusters.
However, no significant effect of EMCB was found for changes in importance in Figure \ref{fig:boxplots}b (see also additional analyses in Supplement \hyperref[app:extended_model]{D}).
This rejects \ref{rq2}a and contradicts earlier findings that show the most pronounced influence of search for people who are already ethically engaged (\citep{santos2024,orourke2016}; see Section \ref{sec_introduction}).

\begin{figure*}[tb]
    \centering
\begin{minipage}{0.45\textwidth}
        \centering
        \begin{tikzpicture}[x=1pt,y=1pt]
\definecolor{fillColor}{RGB}{255,255,255}
\path[use as bounding box,fill=fillColor,fill opacity=0.00] (0,0) rectangle (202.36,108.41);
\begin{scope}
\path[clip] (  0.00,  0.00) rectangle (202.36,108.41);

\path[] (  0.00,  0.00) rectangle (202.36,108.41);
\end{scope}
\begin{scope}
\path[clip] ( 24.26, 27.33) rectangle (196.86,102.90);

\path[] ( 24.26, 27.33) rectangle (196.86,102.91);
\definecolor{drawColor}{RGB}{211,211,211}

\path[draw=drawColor,line width= 0.6pt,line join=round] ( 24.26, 30.77) --
	(196.86, 30.77);

\path[draw=drawColor,line width= 0.6pt,line join=round] ( 24.26, 47.94) --
	(196.86, 47.94);

\path[draw=drawColor,line width= 0.6pt,line join=round] ( 24.26, 65.12) --
	(196.86, 65.12);

\path[draw=drawColor,line width= 0.6pt,line join=round] ( 24.26, 82.29) --
	(196.86, 82.29);

\path[draw=drawColor,line width= 0.6pt,line join=round] ( 24.26, 99.47) --
	(196.86, 99.47);

\path[draw=drawColor,line width= 0.6pt,line join=round] ( 56.62, 27.33) --
	( 56.62,102.90);

\path[draw=drawColor,line width= 0.6pt,line join=round] (110.56, 27.33) --
	(110.56,102.90);

\path[draw=drawColor,line width= 0.6pt,line join=round] (164.49, 27.33) --
	(164.49,102.90);
\definecolor{fillColor}{RGB}{255,0,0}

\path[fill=fillColor] ( 56.62, 82.29) circle (  1.96);

\path[fill=fillColor] ( 56.62, 82.29) circle (  1.96);

\path[fill=fillColor] ( 56.62, 82.29) circle (  1.96);

\path[fill=fillColor] ( 56.62, 82.29) circle (  1.96);

\path[fill=fillColor] ( 56.62, 99.47) circle (  1.96);

\path[fill=fillColor] ( 56.62, 82.29) circle (  1.96);
\definecolor{drawColor}{RGB}{0,0,0}

\path[draw=drawColor,line width= 0.6pt,line join=round] ( 56.62, 47.94) -- ( 56.62, 65.12);

\path[draw=drawColor,line width= 0.6pt,line join=round] ( 56.62, 30.77) -- ( 56.62, 30.77);
\definecolor{fillColor}{RGB}{173,216,230}

\path[draw=drawColor,line width= 0.6pt,fill=fillColor] ( 36.40, 47.94) --
	( 36.40, 30.77) --
	( 76.85, 30.77) --
	( 76.85, 47.94) --
	( 36.40, 47.94) --
	cycle;

\path[draw=drawColor,line width= 1.1pt] ( 36.40, 30.77) -- ( 76.85, 30.77);
\definecolor{fillColor}{RGB}{255,0,0}

\path[fill=fillColor] (110.56, 99.47) circle (  1.96);

\path[fill=fillColor] (110.56, 99.47) circle (  1.96);

\path[fill=fillColor] (110.56, 99.47) circle (  1.96);

\path[fill=fillColor] (110.56, 99.47) circle (  1.96);

\path[fill=fillColor] (110.56, 99.47) circle (  1.96);

\path[draw=drawColor,line width= 0.6pt,line join=round] (110.56, 65.12) -- (110.56, 82.29);

\path[draw=drawColor,line width= 0.6pt,line join=round] (110.56, 47.94) -- (110.56, 30.77);
\definecolor{fillColor}{RGB}{173,216,230}

\path[draw=drawColor,line width= 0.6pt,fill=fillColor] ( 90.33, 65.12) --
	( 90.33, 47.94) --
	(130.79, 47.94) --
	(130.79, 65.12) --
	( 90.33, 65.12) --
	cycle;

\path[draw=drawColor,line width= 1.1pt] ( 90.33, 47.94) -- (130.79, 47.94);
\definecolor{fillColor}{RGB}{255,0,0}

\path[fill=fillColor] (164.49, 30.77) circle (  1.96);

\path[fill=fillColor] (164.49, 30.77) circle (  1.96);

\path[fill=fillColor] (164.49, 30.77) circle (  1.96);

\path[fill=fillColor] (164.49, 30.77) circle (  1.96);

\path[fill=fillColor] (164.49, 30.77) circle (  1.96);

\path[fill=fillColor] (164.49, 30.77) circle (  1.96);

\path[draw=drawColor,line width= 0.6pt,line join=round] (164.49, 82.29) -- (164.49, 99.47);

\path[draw=drawColor,line width= 0.6pt,line join=round] (164.49, 65.12) -- (164.49, 47.94);
\definecolor{fillColor}{RGB}{173,216,230}

\path[draw=drawColor,line width= 0.6pt,fill=fillColor] (144.27, 82.29) --
	(144.27, 65.12) --
	(184.72, 65.12) --
	(184.72, 82.29) --
	(144.27, 82.29) --
	cycle;

\path[draw=drawColor,line width= 1.1pt] (144.27, 82.29) -- (184.72, 82.29);
\end{scope}
\begin{scope}
\path[clip] (  0.00,  0.00) rectangle (202.36,108.41);
\definecolor{drawColor}{RGB}{0,0,0}

\node[text=drawColor,anchor=base east,inner sep=0pt, outer sep=0pt, scale=  0.80] at ( 19.31, 28.01) {1};

\node[text=drawColor,anchor=base east,inner sep=0pt, outer sep=0pt, scale=  0.80] at ( 19.31, 45.19) {2};

\node[text=drawColor,anchor=base east,inner sep=0pt, outer sep=0pt, scale=  0.80] at ( 19.31, 62.36) {3};

\node[text=drawColor,anchor=base east,inner sep=0pt, outer sep=0pt, scale=  0.80] at ( 19.31, 79.54) {4};

\node[text=drawColor,anchor=base east,inner sep=0pt, outer sep=0pt, scale=  0.80] at ( 19.31, 96.71) {5};
\end{scope}
\begin{scope}
\path[clip] (  0.00,  0.00) rectangle (202.36,108.41);
\definecolor{drawColor}{gray}{0.20}

\path[draw=drawColor,line width= 0.6pt,line join=round] ( 21.51, 30.77) --
	( 24.26, 30.77);

\path[draw=drawColor,line width= 0.6pt,line join=round] ( 21.51, 47.94) --
	( 24.26, 47.94);

\path[draw=drawColor,line width= 0.6pt,line join=round] ( 21.51, 65.12) --
	( 24.26, 65.12);

\path[draw=drawColor,line width= 0.6pt,line join=round] ( 21.51, 82.29) --
	( 24.26, 82.29);

\path[draw=drawColor,line width= 0.6pt,line join=round] ( 21.51, 99.47) --
	( 24.26, 99.47);
\end{scope}
\begin{scope}
\path[clip] (  0.00,  0.00) rectangle (202.36,108.41);
\definecolor{drawColor}{gray}{0.20}

\path[draw=drawColor,line width= 0.6pt,line join=round] ( 56.62, 24.58) --
	( 56.62, 27.33);

\path[draw=drawColor,line width= 0.6pt,line join=round] (110.56, 24.58) --
	(110.56, 27.33);

\path[draw=drawColor,line width= 0.6pt,line join=round] (164.49, 24.58) --
	(164.49, 27.33);
\end{scope}
\begin{scope}
\path[clip] (  0.00,  0.00) rectangle (202.36,108.41);
\definecolor{drawColor}{RGB}{0,0,0}

\node[text=drawColor,anchor=base,inner sep=0pt, outer sep=0pt, scale=  0.80] at ( 56.62, 16.87) {Low (2.03)};

\node[text=drawColor,anchor=base,inner sep=0pt, outer sep=0pt, scale=  0.80] at (110.56, 16.87) {Medium (3.15)};

\node[text=drawColor,anchor=base,inner sep=0pt, outer sep=0pt, scale=  0.80] at (164.49, 16.87) {High (4.2)};
\end{scope}
\begin{scope}
\path[clip] (  0.00,  0.00) rectangle (202.36,108.41);
\definecolor{drawColor}{RGB}{0,0,0}

\node[text=drawColor,anchor=base,inner sep=0pt, outer sep=0pt, scale=  0.80] at (110.56,  7.06) {EMCB cluster};
\end{scope}
\begin{scope}
\path[clip] (  0.00,  0.00) rectangle (202.36,108.41);
\definecolor{drawColor}{RGB}{0,0,0}

\node[text=drawColor,rotate= 90.00,anchor=base,inner sep=0pt, outer sep=0pt, scale=  0.80] at ( 11.01, 65.12) {Importance (prior)};
\end{scope}
\end{tikzpicture}
         \caption*{(a)}
    \end{minipage}
    \hfill
\begin{minipage}{0.45\textwidth}
        \centering
        \begin{tikzpicture}[x=1pt,y=1pt]
\definecolor{fillColor}{RGB}{255,255,255}
\path[use as bounding box,fill=fillColor,fill opacity=0.00] (0,0) rectangle (202.36,108.41);
\begin{scope}
\path[clip] (  0.00,  0.00) rectangle (202.36,108.41);

\path[] (  0.00,  0.00) rectangle (202.36,108.41);
\end{scope}
\begin{scope}
\path[clip] ( 26.93, 27.33) rectangle (196.86,102.90);

\path[] ( 26.93, 27.33) rectangle (196.86,102.91);
\definecolor{drawColor}{RGB}{211,211,211}

\path[draw=drawColor,line width= 0.6pt,line join=round] ( 26.93, 30.77) --
	(196.86, 30.77);

\path[draw=drawColor,line width= 0.6pt,line join=round] ( 26.93, 50.40) --
	(196.86, 50.40);

\path[draw=drawColor,line width= 0.6pt,line join=round] ( 26.93, 70.02) --
	(196.86, 70.02);

\path[draw=drawColor,line width= 0.6pt,line join=round] ( 26.93, 89.65) --
	(196.86, 89.65);

\path[draw=drawColor,line width= 0.6pt,line join=round] ( 58.79, 27.33) --
	( 58.79,102.90);

\path[draw=drawColor,line width= 0.6pt,line join=round] (111.89, 27.33) --
	(111.89,102.90);

\path[draw=drawColor,line width= 0.6pt,line join=round] (164.99, 27.33) --
	(164.99,102.90);
\definecolor{fillColor}{RGB}{255,0,0}

\path[fill=fillColor] ( 58.79, 50.40) circle (  1.96);

\path[fill=fillColor] ( 58.79, 50.40) circle (  1.96);

\path[fill=fillColor] ( 58.79, 99.47) circle (  1.96);

\path[fill=fillColor] ( 58.79, 50.40) circle (  1.96);

\path[fill=fillColor] ( 58.79, 40.58) circle (  1.96);

\path[fill=fillColor] ( 58.79, 50.40) circle (  1.96);
\definecolor{drawColor}{RGB}{0,0,0}

\path[draw=drawColor,line width= 0.6pt,line join=round] ( 58.79, 79.84) -- ( 58.79, 89.65);

\path[draw=drawColor,line width= 0.6pt,line join=round] ( 58.79, 70.02) -- ( 58.79, 60.21);
\definecolor{fillColor}{RGB}{173,216,230}

\path[draw=drawColor,line width= 0.6pt,fill=fillColor] ( 38.88, 79.84) --
	( 38.88, 70.02) --
	( 78.70, 70.02) --
	( 78.70, 79.84) --
	( 38.88, 79.84) --
	cycle;

\path[draw=drawColor,line width= 1.1pt] ( 38.88, 70.02) -- ( 78.70, 70.02);
\definecolor{fillColor}{RGB}{255,0,0}

\path[fill=fillColor] (111.89, 99.47) circle (  1.96);

\path[fill=fillColor] (111.89, 99.47) circle (  1.96);

\path[fill=fillColor] (111.89, 99.47) circle (  1.96);

\path[fill=fillColor] (111.89, 50.40) circle (  1.96);

\path[fill=fillColor] (111.89, 50.40) circle (  1.96);

\path[fill=fillColor] (111.89, 30.77) circle (  1.96);

\path[fill=fillColor] (111.89, 99.47) circle (  1.96);

\path[draw=drawColor,line width= 0.6pt,line join=round] (111.89, 79.84) -- (111.89, 89.65);

\path[draw=drawColor,line width= 0.6pt,line join=round] (111.89, 70.02) -- (111.89, 60.21);
\definecolor{fillColor}{RGB}{173,216,230}

\path[draw=drawColor,line width= 0.6pt,fill=fillColor] ( 91.98, 79.84) --
	( 91.98, 70.02) --
	(131.81, 70.02) --
	(131.81, 79.84) --
	( 91.98, 79.84) --
	cycle;

\path[draw=drawColor,line width= 1.1pt] ( 91.98, 70.02) -- (131.81, 70.02);
\definecolor{fillColor}{RGB}{255,0,0}

\path[fill=fillColor] (164.99, 99.47) circle (  1.96);

\path[fill=fillColor] (164.99, 40.58) circle (  1.96);

\path[fill=fillColor] (164.99, 50.40) circle (  1.96);

\path[fill=fillColor] (164.99, 99.47) circle (  1.96);

\path[draw=drawColor,line width= 0.6pt,line join=round] (164.99, 79.84) -- (164.99, 89.65);

\path[draw=drawColor,line width= 0.6pt,line join=round] (164.99, 70.02) -- (164.99, 60.21);
\definecolor{fillColor}{RGB}{173,216,230}

\path[draw=drawColor,line width= 0.6pt,fill=fillColor] (145.08, 79.84) --
	(145.08, 70.02) --
	(184.91, 70.02) --
	(184.91, 79.84) --
	(145.08, 79.84) --
	cycle;

\path[draw=drawColor,line width= 1.1pt] (145.08, 70.02) -- (184.91, 70.02);
\end{scope}
\begin{scope}
\path[clip] (  0.00,  0.00) rectangle (202.36,108.41);
\definecolor{drawColor}{RGB}{0,0,0}

\node[text=drawColor,anchor=base east,inner sep=0pt, outer sep=0pt, scale=  0.80] at ( 21.98, 28.01) {-4};

\node[text=drawColor,anchor=base east,inner sep=0pt, outer sep=0pt, scale=  0.80] at ( 21.98, 47.64) {-2};

\node[text=drawColor,anchor=base east,inner sep=0pt, outer sep=0pt, scale=  0.80] at ( 21.98, 67.27) {0};

\node[text=drawColor,anchor=base east,inner sep=0pt, outer sep=0pt, scale=  0.80] at ( 21.98, 86.90) {2};
\end{scope}
\begin{scope}
\path[clip] (  0.00,  0.00) rectangle (202.36,108.41);
\definecolor{drawColor}{gray}{0.20}

\path[draw=drawColor,line width= 0.6pt,line join=round] ( 24.18, 30.77) --
	( 26.93, 30.77);

\path[draw=drawColor,line width= 0.6pt,line join=round] ( 24.18, 50.40) --
	( 26.93, 50.40);

\path[draw=drawColor,line width= 0.6pt,line join=round] ( 24.18, 70.02) --
	( 26.93, 70.02);

\path[draw=drawColor,line width= 0.6pt,line join=round] ( 24.18, 89.65) --
	( 26.93, 89.65);
\end{scope}
\begin{scope}
\path[clip] (  0.00,  0.00) rectangle (202.36,108.41);
\definecolor{drawColor}{gray}{0.20}

\path[draw=drawColor,line width= 0.6pt,line join=round] ( 58.79, 24.58) --
	( 58.79, 27.33);

\path[draw=drawColor,line width= 0.6pt,line join=round] (111.89, 24.58) --
	(111.89, 27.33);

\path[draw=drawColor,line width= 0.6pt,line join=round] (164.99, 24.58) --
	(164.99, 27.33);
\end{scope}
\begin{scope}
\path[clip] (  0.00,  0.00) rectangle (202.36,108.41);
\definecolor{drawColor}{RGB}{0,0,0}

\node[text=drawColor,anchor=base,inner sep=0pt, outer sep=0pt, scale=  0.80] at ( 58.79, 16.87) {Low (2.03)};

\node[text=drawColor,anchor=base,inner sep=0pt, outer sep=0pt, scale=  0.80] at (111.89, 16.87) {Medium (3.15)};

\node[text=drawColor,anchor=base,inner sep=0pt, outer sep=0pt, scale=  0.80] at (164.99, 16.87) {High (4.2)};
\end{scope}
\begin{scope}
\path[clip] (  0.00,  0.00) rectangle (202.36,108.41);
\definecolor{drawColor}{RGB}{0,0,0}

\node[text=drawColor,anchor=base,inner sep=0pt, outer sep=0pt, scale=  0.80] at (111.89,  7.06) {EMCB cluster};
\end{scope}
\begin{scope}
\path[clip] (  0.00,  0.00) rectangle (202.36,108.41);
\definecolor{drawColor}{RGB}{0,0,0}

\node[text=drawColor,rotate= 90.00,anchor=base,inner sep=0pt, outer sep=0pt, scale=  0.80] at ( 11.01, 65.12) {Importance change};
\end{scope}
\end{tikzpicture}
         \caption*{(b)}
    \end{minipage}

    \caption{
        Box plots illustrating the relationship between EMCB cluster and Importance ratings.
        {\normalfont (a) \emph{Importance (prior)} as a function of \emph{EMCB cluster}, showing a significant ($p < .001$) and large ($\eta^2 = 0.33$) effect of cluster on ratings, $F(2, 308) = 76.06$.
Post-hoc pairwise comparisons using Tukey's HSD test were significant for all comparisons ($p < .001$).
        (b) \emph{Importance change} as a function of \emph{EMCB cluster}, showing no statistically significant effect, $F(2, 308) = 1.93$, $p = .147$, $\eta^2 = 0.01$.
        Post-hoc pairwise comparisons showed no significant differences between the clusters for \emph{Importance change} ($.126 < p < .603$).
    }}
    \label{fig:boxplots}

\end{figure*}

\begin{table*}[tb]
    \centering
    \begin{tabular}{lrrrrrrr}
\toprule 
   & \multicolumn{2}{c}{\textbf{Query Count}} & \multicolumn{2}{c}{\textbf{URL Count}} & \multicolumn{2}{c}{\textbf{Duration (msec)}}  & \textbf{Model} \\ 
 \textbf{Variable} & $\beta$ & t(296) & $\beta$ & t(296) & $\beta$ & t(296) & F(3,296)  \\ 
 \toprule 
  \multicolumn{6}{l}{\textbf{Importance Valuations}} \\ 
\hspace{2mm}\textbf{Prior} & -0.20 & $-3.14^{**}$ & 0.01 & $0.20^{}$ & 0.24 & $3.83^{***}$ & $7.39^{***}$ \\ 
  \hspace{2mm}\textbf{Change} & -0.08 & $-1.18^{}$ & 0.08 & $1.14^{}$ & -0.05 & $-0.75^{}$ & $0.77^{}$ \\ 
   \multicolumn{6}{l}{\textbf{Search Process Appraisals}} \\ 
\hspace{2mm}\textbf{Sense-Making (Ease)} & -0.38 & $-6.10^{***}$ & 0.20 & $2.98^{**}$ & 0.18 & $3.02^{**}$ & $15.04^{***}$ \\ 
  \hspace{2mm}\textbf{Decision-Making (Recognition)} & -0.11 & $-1.66^{}$ & 0.08 & $1.11^{}$ & 0.02 & $0.30^{}$ & $1.02^{}$ \\ 
   \bottomrule
\end{tabular}
 \caption{
        Behavioural metrics regressed on subjective variables.        
        {\normalfont 
        Search queries and page opens were tracked within the search interface; duration measured from first to last action \citep{vandersluis2025ipm}.
        The table reports standardized coefficients ($\beta$) and t-values per behavioural metric, as well as an overall model fit ($F$-statistic), for each subjective variable. Significant effects were noted for \emph{prior importance} valuation, but not for \emph{importance change}; and for the \emph{Search \& evaluation ease} cluster, but not for the \emph{Problem recognition} cluster (see Figure \ref{fig:lfs_clustering} for cluster definitions). Significance levels: $^{*}$ $p<.05$, $^{**}$ $p<.01$, $^{***}$ $p<.001$.
    }}
    \label{tab:behavior}
\end{table*}

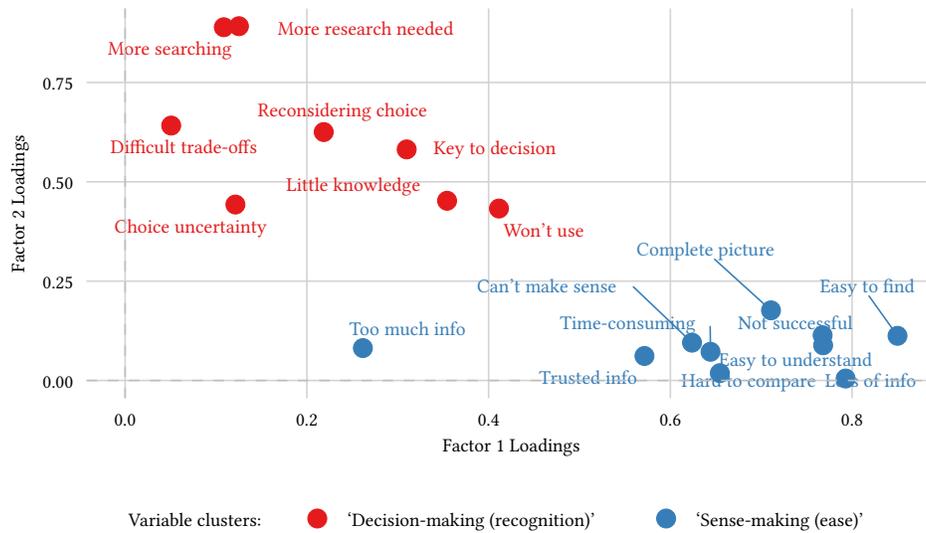
\begin{figure*}[ht]
    \centering
    \begin{tikzpicture}[x=1pt,y=1pt]
\definecolor{fillColor}{RGB}{255,255,255}
\path[use as bounding box,fill=fillColor,fill opacity=0.00] (0,0) rectangle (361.35,216.81);
\begin{scope}
\path[clip] (  0.00,  0.00) rectangle (361.35,216.81);

\path[] (  0.00,  0.00) rectangle (361.35,216.81);
\end{scope}
\begin{scope}
\path[clip] ( 34.48, 63.78) rectangle (355.85,211.31);

\path[] ( 34.48, 63.78) rectangle (355.85,211.31);
\definecolor{drawColor}{RGB}{211,211,211}

\path[draw=drawColor,line width= 0.6pt,line join=round] ( 34.48, 70.49) --
	(355.85, 70.49);

\path[draw=drawColor,line width= 0.6pt,line join=round] ( 34.48,108.09) --
	(355.85,108.09);

\path[draw=drawColor,line width= 0.6pt,line join=round] ( 34.48,145.69) --
	(355.85,145.69);

\path[draw=drawColor,line width= 0.6pt,line join=round] ( 34.48,183.30) --
	(355.85,183.30);

\path[draw=drawColor,line width= 0.6pt,line join=round] ( 49.09, 63.78) --
	( 49.09,211.31);

\path[draw=drawColor,line width= 0.6pt,line join=round] (117.84, 63.78) --
	(117.84,211.31);

\path[draw=drawColor,line width= 0.6pt,line join=round] (186.58, 63.78) --
	(186.58,211.31);

\path[draw=drawColor,line width= 0.6pt,line join=round] (255.33, 63.78) --
	(255.33,211.31);

\path[draw=drawColor,line width= 0.6pt,line join=round] (324.07, 63.78) --
	(324.07,211.31);
\definecolor{drawColor}{RGB}{228,26,28}
\definecolor{fillColor}{RGB}{228,26,28}

\path[draw=drawColor,line width= 0.4pt,line join=round,line cap=round,fill=fillColor] ( 86.42,204.21) circle (  3.57);

\path[draw=drawColor,line width= 0.4pt,line join=round,line cap=round,fill=fillColor] (190.49,135.58) circle (  3.57);
\definecolor{drawColor}{RGB}{55,126,184}
\definecolor{fillColor}{RGB}{55,126,184}

\path[draw=drawColor,line width= 0.4pt,line join=round,line cap=round,fill=fillColor] (341.24, 87.39) circle (  3.57);

\path[draw=drawColor,line width= 0.4pt,line join=round,line cap=round,fill=fillColor] (321.62, 71.23) circle (  3.57);

\path[draw=drawColor,line width= 0.4pt,line join=round,line cap=round,fill=fillColor] (245.53, 79.76) circle (  3.57);

\path[draw=drawColor,line width= 0.4pt,line join=round,line cap=round,fill=fillColor] (263.51, 84.80) circle (  3.57);

\path[draw=drawColor,line width= 0.4pt,line join=round,line cap=round,fill=fillColor] (274.05, 73.26) circle (  3.57);
\definecolor{drawColor}{RGB}{228,26,28}
\definecolor{fillColor}{RGB}{228,26,28}

\path[draw=drawColor,line width= 0.4pt,line join=round,line cap=round,fill=fillColor] ( 90.81,137.09) circle (  3.57);

\path[draw=drawColor,line width= 0.4pt,line join=round,line cap=round,fill=fillColor] ( 92.14,204.60) circle (  3.57);

\path[draw=drawColor,line width= 0.4pt,line join=round,line cap=round,fill=fillColor] (170.87,138.52) circle (  3.57);

\path[draw=drawColor,line width= 0.4pt,line join=round,line cap=round,fill=fillColor] (155.59,157.96) circle (  3.57);
\definecolor{drawColor}{RGB}{55,126,184}
\definecolor{fillColor}{RGB}{55,126,184}

\path[draw=drawColor,line width= 0.4pt,line join=round,line cap=round,fill=fillColor] (270.51, 81.35) circle (  3.57);

\path[draw=drawColor,line width= 0.4pt,line join=round,line cap=round,fill=fillColor] (312.92, 87.61) circle (  3.57);

\path[draw=drawColor,line width= 0.4pt,line join=round,line cap=round,fill=fillColor] (313.11, 83.81) circle (  3.57);

\path[draw=drawColor,line width= 0.4pt,line join=round,line cap=round,fill=fillColor] (293.39, 97.03) circle (  3.57);

\path[draw=drawColor,line width= 0.4pt,line join=round,line cap=round,fill=fillColor] (139.02, 82.75) circle (  3.57);
\definecolor{drawColor}{RGB}{228,26,28}
\definecolor{fillColor}{RGB}{228,26,28}

\path[draw=drawColor,line width= 0.4pt,line join=round,line cap=round,fill=fillColor] ( 66.51,166.98) circle (  3.57);

\path[draw=drawColor,line width= 0.4pt,line join=round,line cap=round,fill=fillColor] (124.21,164.52) circle (  3.57);
\definecolor{drawColor}{RGB}{55,126,184}

\path[draw=drawColor,line width= 0.6pt,line join=round,line cap=round] (330.46,102.47) -- (339.91, 89.25);

\path[draw=drawColor,line width= 0.6pt,line join=round,line cap=round] (241.34,105.93) -- (261.66, 86.57);

\path[draw=drawColor,line width= 0.6pt,line join=round,line cap=round] (270.38, 90.89) -- (270.49, 83.32);

\path[draw=drawColor,line width= 0.6pt,line join=round,line cap=round] (272.07,116.37) -- (291.46, 98.78);
\definecolor{drawColor}{RGB}{228,26,28}

\node[text=drawColor,anchor=base,inner sep=0pt, outer sep=0pt, scale=  0.82] at ( 66.00,193.62) {More searching};

\node[text=drawColor,anchor=base,inner sep=0pt, outer sep=0pt, scale=  0.82] at (207.42,124.99) {Won't use};
\definecolor{drawColor}{RGB}{55,126,184}

\node[text=drawColor,anchor=base,inner sep=0pt, outer sep=0pt, scale=  0.82] at (329.82,103.98) {Easy to find};

\node[text=drawColor,anchor=base,inner sep=0pt, outer sep=0pt, scale=  0.82] at (331.07, 68.00) {Lots of info};

\node[text=drawColor,anchor=base,inner sep=0pt, outer sep=0pt, scale=  0.82] at (224.24, 69.18) {Trusted info};

\node[text=drawColor,anchor=base,inner sep=0pt, outer sep=0pt, scale=  0.82] at (208.62,103.80) {Can't make sense};

\node[text=drawColor,anchor=base,inner sep=0pt, outer sep=0pt, scale=  0.82] at (284.93, 68.00) {Hard to compare};
\definecolor{drawColor}{RGB}{228,26,28}

\node[text=drawColor,anchor=base,inner sep=0pt, outer sep=0pt, scale=  0.82] at ( 73.92,126.52) {Choice uncertainty};

\node[text=drawColor,anchor=base,inner sep=0pt, outer sep=0pt, scale=  0.82] at (140.07,201.45) {More research needed};

\node[text=drawColor,anchor=base,inner sep=0pt, outer sep=0pt, scale=  0.82] at (135.46,142.25) {Little knowledge};

\node[text=drawColor,anchor=base,inner sep=0pt, outer sep=0pt, scale=  0.82] at (188.96,156.33) {Key to decision};
\definecolor{drawColor}{RGB}{55,126,184}

\node[text=drawColor,anchor=base,inner sep=0pt, outer sep=0pt, scale=  0.82] at (239.27, 89.74) {Time-consuming};

\node[text=drawColor,anchor=base,inner sep=0pt, outer sep=0pt, scale=  0.82] at (302.63, 89.93) {Not successful};

\node[text=drawColor,anchor=base,inner sep=0pt, outer sep=0pt, scale=  0.82] at (302.65, 76.15) {Easy to understand};

\node[text=drawColor,anchor=base,inner sep=0pt, outer sep=0pt, scale=  0.82] at (268.67,117.88) {Complete picture};

\node[text=drawColor,anchor=base,inner sep=0pt, outer sep=0pt, scale=  0.82] at (155.93, 87.69) {Too much info};
\definecolor{drawColor}{RGB}{228,26,28}

\node[text=drawColor,anchor=base,inner sep=0pt, outer sep=0pt, scale=  0.82] at ( 71.33,156.36) {Difficult trade-offs};

\node[text=drawColor,anchor=base,inner sep=0pt, outer sep=0pt, scale=  0.82] at (131.35,170.42) {Reconsidering choice};
\definecolor{drawColor}{RGB}{190,190,190}

\path[draw=drawColor,line width= 0.6pt,dash pattern=on 4pt off 4pt ,line join=round] ( 34.48, 70.49) -- (355.85, 70.49);

\path[draw=drawColor,line width= 0.6pt,dash pattern=on 4pt off 4pt ,line join=round] ( 49.09, 63.78) -- ( 49.09,211.31);
\end{scope}
\begin{scope}
\path[clip] (  0.00,  0.00) rectangle (361.35,216.81);
\definecolor{drawColor}{RGB}{0,0,0}

\node[text=drawColor,anchor=base east,inner sep=0pt, outer sep=0pt, scale=  0.80] at ( 29.53, 67.73) {0.00};

\node[text=drawColor,anchor=base east,inner sep=0pt, outer sep=0pt, scale=  0.80] at ( 29.53,105.34) {0.25};

\node[text=drawColor,anchor=base east,inner sep=0pt, outer sep=0pt, scale=  0.80] at ( 29.53,142.94) {0.50};

\node[text=drawColor,anchor=base east,inner sep=0pt, outer sep=0pt, scale=  0.80] at ( 29.53,180.54) {0.75};
\end{scope}
\begin{scope}
\path[clip] (  0.00,  0.00) rectangle (361.35,216.81);
\definecolor{drawColor}{RGB}{0,0,0}

\node[text=drawColor,anchor=base,inner sep=0pt, outer sep=0pt, scale=  0.80] at ( 49.09, 53.32) {0.0};

\node[text=drawColor,anchor=base,inner sep=0pt, outer sep=0pt, scale=  0.80] at (117.84, 53.32) {0.2};

\node[text=drawColor,anchor=base,inner sep=0pt, outer sep=0pt, scale=  0.80] at (186.58, 53.32) {0.4};

\node[text=drawColor,anchor=base,inner sep=0pt, outer sep=0pt, scale=  0.80] at (255.33, 53.32) {0.6};

\node[text=drawColor,anchor=base,inner sep=0pt, outer sep=0pt, scale=  0.80] at (324.07, 53.32) {0.8};
\end{scope}
\begin{scope}
\path[clip] (  0.00,  0.00) rectangle (361.35,216.81);
\definecolor{drawColor}{RGB}{0,0,0}

\node[text=drawColor,anchor=base,inner sep=0pt, outer sep=0pt, scale=  0.80] at (195.17, 43.51) {Factor 1 Loadings};
\end{scope}
\begin{scope}
\path[clip] (  0.00,  0.00) rectangle (361.35,216.81);
\definecolor{drawColor}{RGB}{0,0,0}

\node[text=drawColor,rotate= 90.00,anchor=base,inner sep=0pt, outer sep=0pt, scale=  0.80] at ( 11.01,137.55) {Factor 2 Loadings};
\end{scope}
\begin{scope}
\path[clip] (  0.00,  0.00) rectangle (361.35,216.81);

\path[] ( 44.93,  5.50) rectangle (345.40, 30.95);
\end{scope}
\begin{scope}
\path[clip] (  0.00,  0.00) rectangle (361.35,216.81);
\definecolor{drawColor}{RGB}{0,0,0}

\node[text=drawColor,anchor=base west,inner sep=0pt, outer sep=0pt, scale=  0.80] at ( 50.43, 15.47) {Variable clusters:};
\end{scope}
\begin{scope}
\path[clip] (  0.00,  0.00) rectangle (361.35,216.81);
\definecolor{drawColor}{RGB}{228,26,28}
\definecolor{fillColor}{RGB}{228,26,28}

\path[draw=drawColor,line width= 0.4pt,line join=round,line cap=round,fill=fillColor] (121.78, 18.23) circle (  3.57);
\end{scope}
\begin{scope}
\path[clip] (  0.00,  0.00) rectangle (361.35,216.81);
\definecolor{drawColor}{RGB}{228,26,28}

\node[text=drawColor,anchor=base,inner sep=0pt, outer sep=0pt, scale=  0.82] at (121.78, 15.41) {a};
\end{scope}
\begin{scope}
\path[clip] (  0.00,  0.00) rectangle (361.35,216.81);
\definecolor{drawColor}{RGB}{55,126,184}
\definecolor{fillColor}{RGB}{55,126,184}

\path[draw=drawColor,line width= 0.4pt,line join=round,line cap=round,fill=fillColor] (253.72, 18.23) circle (  3.57);
\end{scope}
\begin{scope}
\path[clip] (  0.00,  0.00) rectangle (361.35,216.81);
\definecolor{drawColor}{RGB}{55,126,184}

\node[text=drawColor,anchor=base,inner sep=0pt, outer sep=0pt, scale=  0.82] at (253.72, 15.41) {a};
\end{scope}
\begin{scope}
\path[clip] (  0.00,  0.00) rectangle (361.35,216.81);
\definecolor{drawColor}{RGB}{0,0,0}

\node[text=drawColor,anchor=base west,inner sep=0pt, outer sep=0pt, scale=  0.80] at (133.00, 15.47) {`Decision-making (recognition)'};
\end{scope}
\begin{scope}
\path[clip] (  0.00,  0.00) rectangle (361.35,216.81);
\definecolor{drawColor}{RGB}{0,0,0}

\node[text=drawColor,anchor=base west,inner sep=0pt, outer sep=0pt, scale=  0.80] at (264.94, 15.47) {`Sense-making (ease)'};
\end{scope}
\end{tikzpicture}
     \caption{
        Exploratory analysis of variable groupings using EFA followed by KMeans clustering.
        {\normalfont
        Exploratory Factor Analysis (EFA) was performed to extract two factors, with variables plotted based on their absolute factor loadings. 
        KMeans clustering was then applied to group variables, with the optimal two-cluster solution determined through silhouette analysis and the scree plot (Supplement \hyperref[app:variable_clustering]{F}). 
        The clusters are labelled \emph{decision-making (recognition)} (Cronbach's $\alpha = .84$) and \emph{sense-making (ease)} (Cronbach's $\alpha = .88$), reflecting the characteristics of the grouped variables.
    }}
    \label{fig:lfs_clustering}
\end{figure*}

\subsection{Search Process and Importance (Change)}

Table~\ref{tab:behavior} presents regression models linking behavioral indexes to both importance and importance change.
Participants who rated an ethical aspect as more important before searching used fewer queries ($\beta = -0.20$, $p < 0.01$) but spent more time searching ($\beta = 0.24$, $p < 0.001$).
This indicates that higher importance led to more focused, deliberate searches, possibly driven by both familiarity with the aspect and increased interest in the results~\citep[cf.~][]{vandersluis2025ipm}.
However, the table shows no evidence of a relationship between search activity and importance change.
Although searching had an overall effect on importance change, these findings negate \ref{rq2}b that search activity alone is sufficient to drive this change.

To further investigate the influence of the search process, we analysed the subjective questions asked about participants' searches.
Exploratory factor analysis, a standard step for custom-developed questionnaires, was used to assess how the questions, based on the four phases of the consumer search model, clustered together (Figure~\ref{fig:lfs_clustering}).
Two distinct clusters of variables emerged, which we labeled `decision-making' and `sense-making' variables.
The decision-making cluster recognises
ethical aspects as part of purchase decisions, complicating decision making and needing extra information.
Key examples include the need for more research (\emph{More research needed}) and reconsideration of choices (\emph{Reconsidering choice}), as well as uncertainty about selecting a product (\emph{Choice uncertainty}) and the recognition that certain aspects were key to the decision (\emph{Key to decision}).
The sense-making cluster focuses on how participants found and processed information, such as whether they found the information easy to locate (\emph{Easy to find}) or trustworthy (\emph{Trusted info}), and whether they were successful in making sense of it (\emph{Can't make sense}).
These clusters map onto the underlying search process model, with the decision-making cluster relating to awareness (Phase I) of knowledge gaps and recognition (Phase IV) of decision gaps and the sense-making cluster tied to the search (Phase II) and evaluation (Phase III) phases.

Figure~\ref{fig:importancechange} shows how the clusters of subjective variables explain changes in importance valuation.
Both clusters had highly significant main effects, with recognition ($\beta = 0.23$, $p < 0.001$) and ease ($\beta = 0.17$, $p < 0.001$) each playing a key role in predicting importance change. These strong effects stand in contrast to other non-significant relationships found with search activity (Table~\ref{tab:behavior}) or behavioral intentions (Figure~\ref{fig:boxplots}).
These findings rather indicate that the subjective dimensions related to the search process are necessary to explain changes in importance, 
answering \ref{rq2}c.

Figure~\ref{fig:importancechange} (left graph) furthermore reveals that EMCB moderates the relationship between recognition and importance change ($\beta = -0.28$, $p < 0.05$).
For participants with high ethical intentions, recognition shows no effect on importance change in Figure~\ref{fig:importancechange} (left graph), whereas for those with lower EMCB scores, it strengthened the effect of recognition.
Together with Figure~\ref{fig:boxplots}, this demonstrates that while higher EMCB scores increase the initial importance assigned to ethical aspects, lower intentions correspond to a greater potential for change.
This positions decision-making recognition as the 
catalyst that turns (low) intentions into increased importance valuations.

Table~\ref{tab:behavior} presents further analyses linking the search process clusters to search activity. 
While search activity was unrelated to decision-making recognition, it was strongly associated with participants' perceived sense-making. 
Characterized by fewer queries ($\beta = -0.38$, $p < 0.001$), more URLs opened ($\beta = 0.20$, $p < 0.01$), and longer search duration ($\beta = 0.18$, $p < 0.01$), this suggests that participants who found the search process smoother and more effective were more focused and engaged.
Interestingly, this contrasts with the lack of a direct relationship between search activity and importance change in Table~\ref{tab:behavior}. 
For search activity to impact importance change, it appears it must first shape participants' sense of ease and success in understanding. 
Like decision-making recognition, this positions subjective appraisals of the search process as a key mediator between search activity and changes in importance.

\subsection{Qualitative analyses}
The previous sections highlighted the important roles of subjectively 
recognizing and resolving gaps in knowledge and decision-making. 
However, these findings do not clarify what drives individuals to identify these gaps or what enables them to make sense of them, nor how these processes 
influence importance change.
To explore this further, we analysed participants' qualitative comments about their decision-making processes, learnings, and issues during their searches, of which we highlight key examples here.

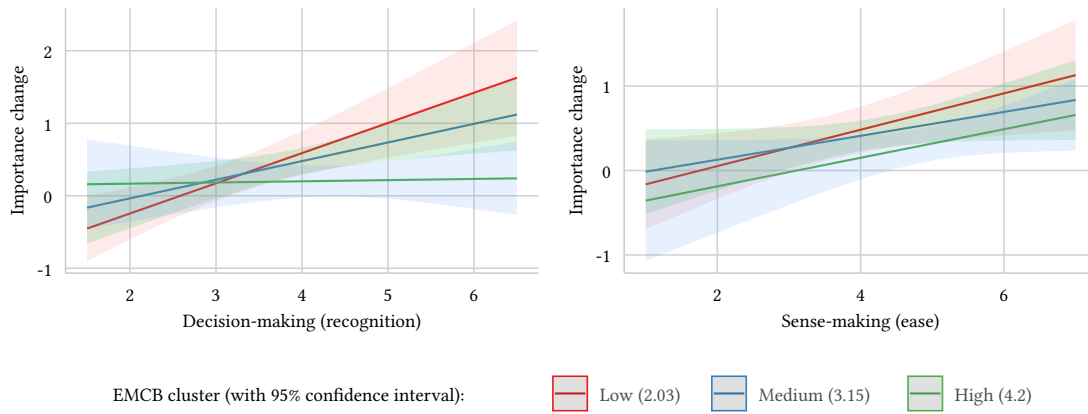
\begin{figure*}[tb]
    \begin{tikzpicture}[x=1pt,y=1pt]
\definecolor{fillColor}{RGB}{255,255,255}
\path[use as bounding box,fill=fillColor,fill opacity=0.00] (0,0) rectangle (433.62,180.67);
\begin{scope}
\path[clip] (  0.00,  0.00) rectangle (433.62,180.67);
\definecolor{drawColor}{RGB}{255,255,255}
\definecolor{fillColor}{RGB}{255,255,255}

\path[draw=drawColor,line width= 0.6pt,line join=round,line cap=round,fill=fillColor] (  0.00,  0.00) rectangle (433.62,180.67);
\end{scope}
\begin{scope}
\path[clip] (  5.50,  5.50) rectangle (216.81,175.17);

\path[] (  5.50,  5.50) rectangle (216.81,175.18);
\end{scope}
\begin{scope}
\path[clip] ( 32.43, 69.28) rectangle (211.31,169.67);

\path[] ( 32.43, 69.28) rectangle (211.31,169.67);
\definecolor{drawColor}{RGB}{211,211,211}

\path[draw=drawColor,line width= 0.6pt,line join=round] ( 32.43, 71.14) --
	(211.31, 71.14);

\path[draw=drawColor,line width= 0.6pt,line join=round] ( 32.43, 98.57) --
	(211.31, 98.57);

\path[draw=drawColor,line width= 0.6pt,line join=round] ( 32.43,126.01) --
	(211.31,126.01);

\path[draw=drawColor,line width= 0.6pt,line join=round] ( 32.43,153.45) --
	(211.31,153.45);

\path[draw=drawColor,line width= 0.6pt,line join=round] ( 56.82, 69.28) --
	( 56.82,169.67);

\path[draw=drawColor,line width= 0.6pt,line join=round] ( 89.35, 69.28) --
	( 89.35,169.67);

\path[draw=drawColor,line width= 0.6pt,line join=round] (121.87, 69.28) --
	(121.87,169.67);

\path[draw=drawColor,line width= 0.6pt,line join=round] (154.39, 69.28) --
	(154.39,169.67);

\path[draw=drawColor,line width= 0.6pt,line join=round] (186.92, 69.28) --
	(186.92,169.67);
\definecolor{drawColor}{RGB}{228,26,28}

\path[draw=drawColor,line width= 0.8pt,line join=round] ( 40.56, 86.16) --
	( 56.82, 91.87) --
	( 73.08, 97.58) --
	( 89.35,103.29) --
	(105.61,109.00) --
	(121.87,114.71) --
	(138.13,120.42) --
	(154.39,126.13) --
	(170.66,131.84) --
	(186.92,137.55) --
	(203.18,143.26);
\definecolor{drawColor}{RGB}{55,126,184}

\path[draw=drawColor,line width= 0.8pt,line join=round] ( 40.56, 94.07) --
	( 56.82, 97.59) --
	( 73.08,101.12) --
	( 89.35,104.64) --
	(105.61,108.17) --
	(121.87,111.69) --
	(138.13,115.22) --
	(154.39,118.74) --
	(170.66,122.27) --
	(186.92,125.80) --
	(203.18,129.32);
\definecolor{drawColor}{RGB}{77,175,74}

\path[draw=drawColor,line width= 0.8pt,line join=round] ( 40.56,102.93) --
	( 56.82,103.15) --
	( 73.08,103.38) --
	( 89.35,103.60) --
	(105.61,103.82) --
	(121.87,104.04) --
	(138.13,104.27) --
	(154.39,104.49) --
	(170.66,104.71) --
	(186.92,104.93) --
	(203.18,105.16);
\definecolor{fillColor}{RGB}{248,118,109}

\path[fill=fillColor,fill opacity=0.15] ( 40.56, 98.48) --
	( 56.82,101.67) --
	( 73.08,105.32) --
	( 89.35,109.88) --
	(105.61,115.82) --
	(121.87,123.03) --
	(138.13,130.99) --
	(154.39,139.31) --
	(170.66,147.82) --
	(186.92,156.43) --
	(203.18,165.11) --
	(203.18,121.41) --
	(186.92,118.67) --
	(170.66,115.87) --
	(154.39,112.96) --
	(138.13,109.86) --
	(121.87,106.39) --
	(105.61,102.19) --
	( 89.35, 96.71) --
	( 73.08, 89.85) --
	( 56.82, 82.08) --
	( 40.56, 73.85) --
	cycle;

\path[] ( 40.56, 98.48) --
	( 56.82,101.67) --
	( 73.08,105.32) --
	( 89.35,109.88) --
	(105.61,115.82) --
	(121.87,123.03) --
	(138.13,130.99) --
	(154.39,139.31) --
	(170.66,147.82) --
	(186.92,156.43) --
	(203.18,165.11);

\path[] (203.18,121.41) --
	(186.92,118.67) --
	(170.66,115.87) --
	(154.39,112.96) --
	(138.13,109.86) --
	(121.87,106.39) --
	(105.61,102.19) --
	( 89.35, 96.71) --
	( 73.08, 89.85) --
	( 56.82, 82.08) --
	( 40.56, 73.85);
\definecolor{fillColor}{RGB}{0,186,56}

\path[fill=fillColor,fill opacity=0.15] ( 40.56,107.72) --
	( 56.82,108.91) --
	( 73.08,110.21) --
	( 89.35,111.73) --
	(105.61,113.71) --
	(121.87,116.60) --
	(138.13,120.74) --
	(154.39,125.80) --
	(170.66,131.33) --
	(186.92,137.08) --
	(203.18,142.94) --
	(203.18,115.70) --
	(186.92,114.51) --
	(170.66,113.21) --
	(154.39,111.69) --
	(138.13,109.70) --
	(121.87,106.79) --
	(105.61,102.63) --
	( 89.35, 97.56) --
	( 73.08, 92.02) --
	( 56.82, 86.27) --
	( 40.56, 80.41) --
	cycle;

\path[] ( 40.56,107.72) --
	( 56.82,108.91) --
	( 73.08,110.21) --
	( 89.35,111.73) --
	(105.61,113.71) --
	(121.87,116.60) --
	(138.13,120.74) --
	(154.39,125.80) --
	(170.66,131.33) --
	(186.92,137.08) --
	(203.18,142.94);

\path[] (203.18,115.70) --
	(186.92,114.51) --
	(170.66,113.21) --
	(154.39,111.69) --
	(138.13,109.70) --
	(121.87,106.79) --
	(105.61,102.63) --
	( 89.35, 97.56) --
	( 73.08, 92.02) --
	( 56.82, 86.27) --
	( 40.56, 80.41);
\definecolor{fillColor}{RGB}{97,156,255}

\path[fill=fillColor,fill opacity=0.15] ( 40.56,119.99) --
	( 56.82,117.55) --
	( 73.08,115.20) --
	( 89.35,113.03) --
	(105.61,111.20) --
	(121.87,110.08) --
	(138.13,110.16) --
	(154.39,111.52) --
	(170.66,113.69) --
	(186.92,116.26) --
	(203.18,119.02) --
	(203.18, 91.29) --
	(186.92, 93.61) --
	(170.66, 95.73) --
	(154.39, 97.45) --
	(138.13, 98.37) --
	(121.87, 98.01) --
	(105.61, 96.44) --
	( 89.35, 94.17) --
	( 73.08, 91.55) --
	( 56.82, 88.76) --
	( 40.56, 85.87) --
	cycle;

\path[] ( 40.56,119.99) --
	( 56.82,117.55) --
	( 73.08,115.20) --
	( 89.35,113.03) --
	(105.61,111.20) --
	(121.87,110.08) --
	(138.13,110.16) --
	(154.39,111.52) --
	(170.66,113.69) --
	(186.92,116.26) --
	(203.18,119.02);

\path[] (203.18, 91.29) --
	(186.92, 93.61) --
	(170.66, 95.73) --
	(154.39, 97.45) --
	(138.13, 98.37) --
	(121.87, 98.01) --
	(105.61, 96.44) --
	( 89.35, 94.17) --
	( 73.08, 91.55) --
	( 56.82, 88.76) --
	( 40.56, 85.87);
\end{scope}
\begin{scope}
\path[clip] (  0.00,  0.00) rectangle (433.62,180.67);
\definecolor{drawColor}{gray}{0.80}

\path[draw=drawColor,line width= 0.6pt,line join=round] ( 32.43, 69.28) --
	( 32.43,169.67);
\end{scope}
\begin{scope}
\path[clip] (  0.00,  0.00) rectangle (433.62,180.67);
\definecolor{drawColor}{RGB}{0,0,0}

\node[text=drawColor,anchor=base east,inner sep=0pt, outer sep=0pt, scale=  0.80] at ( 27.48, 68.38) {-1};

\node[text=drawColor,anchor=base east,inner sep=0pt, outer sep=0pt, scale=  0.80] at ( 27.48, 95.82) {0};

\node[text=drawColor,anchor=base east,inner sep=0pt, outer sep=0pt, scale=  0.80] at ( 27.48,123.26) {1};

\node[text=drawColor,anchor=base east,inner sep=0pt, outer sep=0pt, scale=  0.80] at ( 27.48,150.70) {2};
\end{scope}
\begin{scope}
\path[clip] (  0.00,  0.00) rectangle (433.62,180.67);
\definecolor{drawColor}{gray}{0.80}

\path[draw=drawColor,line width= 0.6pt,line join=round] ( 32.43, 69.28) --
	(211.31, 69.28);
\end{scope}
\begin{scope}
\path[clip] (  0.00,  0.00) rectangle (433.62,180.67);
\definecolor{drawColor}{RGB}{0,0,0}

\node[text=drawColor,anchor=base,inner sep=0pt, outer sep=0pt, scale=  0.80] at ( 56.82, 58.82) {2};

\node[text=drawColor,anchor=base,inner sep=0pt, outer sep=0pt, scale=  0.80] at ( 89.35, 58.82) {3};

\node[text=drawColor,anchor=base,inner sep=0pt, outer sep=0pt, scale=  0.80] at (121.87, 58.82) {4};

\node[text=drawColor,anchor=base,inner sep=0pt, outer sep=0pt, scale=  0.80] at (154.39, 58.82) {5};

\node[text=drawColor,anchor=base,inner sep=0pt, outer sep=0pt, scale=  0.80] at (186.92, 58.82) {6};
\end{scope}
\begin{scope}
\path[clip] (  0.00,  0.00) rectangle (433.62,180.67);
\definecolor{drawColor}{RGB}{0,0,0}

\node[text=drawColor,anchor=base,inner sep=0pt, outer sep=0pt, scale=  0.80] at (121.87, 49.01) {Decision-making (recognition)};
\end{scope}
\begin{scope}
\path[clip] (  0.00,  0.00) rectangle (433.62,180.67);
\definecolor{drawColor}{RGB}{0,0,0}

\node[text=drawColor,rotate= 90.00,anchor=base,inner sep=0pt, outer sep=0pt, scale=  0.80] at ( 16.51,119.48) {Importance change};
\end{scope}
\begin{scope}
\path[clip] (216.81,  5.50) rectangle (428.12,175.17);

\path[] (216.81,  5.50) rectangle (428.12,175.18);
\end{scope}
\begin{scope}
\path[clip] (243.74, 69.28) rectangle (422.62,169.67);

\path[] (243.74, 69.28) rectangle (422.62,169.67);
\definecolor{drawColor}{RGB}{211,211,211}

\path[draw=drawColor,line width= 0.6pt,line join=round] (243.74, 76.16) --
	(422.62, 76.16);

\path[draw=drawColor,line width= 0.6pt,line join=round] (243.74,108.12) --
	(422.62,108.12);

\path[draw=drawColor,line width= 0.6pt,line join=round] (243.74,140.08) --
	(422.62,140.08);

\path[draw=drawColor,line width= 0.6pt,line join=round] (278.97, 69.28) --
	(278.97,169.67);

\path[draw=drawColor,line width= 0.6pt,line join=round] (333.18, 69.28) --
	(333.18,169.67);

\path[draw=drawColor,line width= 0.6pt,line join=round] (387.39, 69.28) --
	(387.39,169.67);
\definecolor{drawColor}{RGB}{228,26,28}

\path[draw=drawColor,line width= 0.8pt,line join=round] (251.87,102.90) --
	(265.42,106.35) --
	(278.97,109.79) --
	(292.53,113.23) --
	(306.08,116.68) --
	(319.63,120.12) --
	(333.18,123.57) --
	(346.73,127.01) --
	(360.28,130.45) --
	(373.83,133.90) --
	(387.39,137.34) --
	(400.94,140.79) --
	(414.49,144.23);
\definecolor{drawColor}{RGB}{55,126,184}

\path[draw=drawColor,line width= 0.8pt,line join=round] (251.87,107.69) --
	(265.42,109.95) --
	(278.97,112.21) --
	(292.53,114.47) --
	(306.08,116.73) --
	(319.63,119.00) --
	(333.18,121.26) --
	(346.73,123.52) --
	(360.28,125.78) --
	(373.83,128.04) --
	(387.39,130.30) --
	(400.94,132.56) --
	(414.49,134.83);
\definecolor{drawColor}{RGB}{77,175,74}

\path[draw=drawColor,line width= 0.8pt,line join=round] (251.87, 96.73) --
	(265.42, 99.43) --
	(278.97,102.14) --
	(292.53,104.84) --
	(306.08,107.54) --
	(319.63,110.24) --
	(333.18,112.95) --
	(346.73,115.65) --
	(360.28,118.35) --
	(373.83,121.05) --
	(387.39,123.75) --
	(400.94,126.46) --
	(414.49,129.16);
\definecolor{fillColor}{RGB}{248,118,109}

\path[fill=fillColor,fill opacity=0.15] (251.87,119.89) --
	(265.42,120.95) --
	(278.97,122.20) --
	(292.53,123.75) --
	(306.08,125.79) --
	(319.63,128.57) --
	(333.18,132.27) --
	(346.73,136.81) --
	(360.28,141.95) --
	(373.83,147.47) --
	(387.39,153.22) --
	(400.94,159.12) --
	(414.49,165.11) --
	(414.49,123.35) --
	(400.94,122.45) --
	(387.39,121.46) --
	(373.83,120.33) --
	(360.28,118.96) --
	(346.73,117.22) --
	(333.18,114.87) --
	(319.63,111.67) --
	(306.08,107.57) --
	(292.53,102.72) --
	(278.97, 97.38) --
	(265.42, 91.74) --
	(251.87, 85.91) --
	cycle;

\path[] (251.87,119.89) --
	(265.42,120.95) --
	(278.97,122.20) --
	(292.53,123.75) --
	(306.08,125.79) --
	(319.63,128.57) --
	(333.18,132.27) --
	(346.73,136.81) --
	(360.28,141.95) --
	(373.83,147.47) --
	(387.39,153.22) --
	(400.94,159.12) --
	(414.49,165.11);

\path[] (414.49,123.35) --
	(400.94,122.45) --
	(387.39,121.46) --
	(373.83,120.33) --
	(360.28,118.96) --
	(346.73,117.22) --
	(333.18,114.87) --
	(319.63,111.67) --
	(306.08,107.57) --
	(292.53,102.72) --
	(278.97, 97.38) --
	(265.42, 91.74) --
	(251.87, 85.91);
\definecolor{fillColor}{RGB}{0,186,56}

\path[fill=fillColor,fill opacity=0.15] (251.87,123.74) --
	(265.42,123.77) --
	(278.97,123.87) --
	(292.53,124.09) --
	(306.08,124.54) --
	(319.63,125.39) --
	(333.18,126.97) --
	(346.73,129.53) --
	(360.28,132.94) --
	(373.83,136.88) --
	(387.39,141.11) --
	(400.94,145.49) --
	(414.49,149.96) --
	(414.49,119.70) --
	(400.94,119.64) --
	(387.39,119.50) --
	(373.83,119.20) --
	(360.28,118.62) --
	(346.73,117.51) --
	(333.18,115.54) --
	(319.63,112.60) --
	(306.08,108.93) --
	(292.53,104.85) --
	(278.97,100.55) --
	(265.42, 96.13) --
	(251.87, 91.63) --
	cycle;

\path[] (251.87,123.74) --
	(265.42,123.77) --
	(278.97,123.87) --
	(292.53,124.09) --
	(306.08,124.54) --
	(319.63,125.39) --
	(333.18,126.97) --
	(346.73,129.53) --
	(360.28,132.94) --
	(373.83,136.88) --
	(387.39,141.11) --
	(400.94,145.49) --
	(414.49,149.96);

\path[] (414.49,119.70) --
	(400.94,119.64) --
	(387.39,119.50) --
	(373.83,119.20) --
	(360.28,118.62) --
	(346.73,117.51) --
	(333.18,115.54) --
	(319.63,112.60) --
	(306.08,108.93) --
	(292.53,104.85) --
	(278.97,100.55) --
	(265.42, 96.13) --
	(251.87, 91.63);
\definecolor{fillColor}{RGB}{97,156,255}

\path[fill=fillColor,fill opacity=0.15] (251.87,119.62) --
	(265.42,119.62) --
	(278.97,119.66) --
	(292.53,119.77) --
	(306.08,119.98) --
	(319.63,120.36) --
	(333.18,121.07) --
	(346.73,122.41) --
	(360.28,124.78) --
	(373.83,128.33) --
	(387.39,132.73) --
	(400.94,137.60) --
	(414.49,142.72) --
	(414.49,115.60) --
	(400.94,115.31) --
	(387.39,114.77) --
	(373.83,113.77) --
	(360.28,111.92) --
	(346.73,108.89) --
	(333.18,104.82) --
	(319.63,100.13) --
	(306.08, 95.11) --
	(292.53, 89.91) --
	(278.97, 84.61) --
	(265.42, 79.25) --
	(251.87, 73.85) --
	cycle;

\path[] (251.87,119.62) --
	(265.42,119.62) --
	(278.97,119.66) --
	(292.53,119.77) --
	(306.08,119.98) --
	(319.63,120.36) --
	(333.18,121.07) --
	(346.73,122.41) --
	(360.28,124.78) --
	(373.83,128.33) --
	(387.39,132.73) --
	(400.94,137.60) --
	(414.49,142.72);

\path[] (414.49,115.60) --
	(400.94,115.31) --
	(387.39,114.77) --
	(373.83,113.77) --
	(360.28,111.92) --
	(346.73,108.89) --
	(333.18,104.82) --
	(319.63,100.13) --
	(306.08, 95.11) --
	(292.53, 89.91) --
	(278.97, 84.61) --
	(265.42, 79.25) --
	(251.87, 73.85);
\end{scope}
\begin{scope}
\path[clip] (  0.00,  0.00) rectangle (433.62,180.67);
\definecolor{drawColor}{gray}{0.80}

\path[draw=drawColor,line width= 0.6pt,line join=round] (243.74, 69.28) --
	(243.74,169.67);
\end{scope}
\begin{scope}
\path[clip] (  0.00,  0.00) rectangle (433.62,180.67);
\definecolor{drawColor}{RGB}{0,0,0}

\node[text=drawColor,anchor=base east,inner sep=0pt, outer sep=0pt, scale=  0.80] at (238.79, 73.40) {-1};

\node[text=drawColor,anchor=base east,inner sep=0pt, outer sep=0pt, scale=  0.80] at (238.79,105.36) {0};

\node[text=drawColor,anchor=base east,inner sep=0pt, outer sep=0pt, scale=  0.80] at (238.79,137.32) {1};
\end{scope}
\begin{scope}
\path[clip] (  0.00,  0.00) rectangle (433.62,180.67);
\definecolor{drawColor}{gray}{0.80}

\path[draw=drawColor,line width= 0.6pt,line join=round] (243.74, 69.28) --
	(422.62, 69.28);
\end{scope}
\begin{scope}
\path[clip] (  0.00,  0.00) rectangle (433.62,180.67);
\definecolor{drawColor}{RGB}{0,0,0}

\node[text=drawColor,anchor=base,inner sep=0pt, outer sep=0pt, scale=  0.80] at (278.97, 58.82) {2};

\node[text=drawColor,anchor=base,inner sep=0pt, outer sep=0pt, scale=  0.80] at (333.18, 58.82) {4};

\node[text=drawColor,anchor=base,inner sep=0pt, outer sep=0pt, scale=  0.80] at (387.39, 58.82) {6};
\end{scope}
\begin{scope}
\path[clip] (  0.00,  0.00) rectangle (433.62,180.67);
\definecolor{drawColor}{RGB}{0,0,0}

\node[text=drawColor,anchor=base,inner sep=0pt, outer sep=0pt, scale=  0.80] at (333.18, 49.01) {Sense-making (ease)};
\end{scope}
\begin{scope}
\path[clip] (  0.00,  0.00) rectangle (433.62,180.67);
\definecolor{drawColor}{RGB}{0,0,0}

\node[text=drawColor,rotate= 90.00,anchor=base,inner sep=0pt, outer sep=0pt, scale=  0.80] at (227.82,119.48) {Importance change};
\end{scope}
\begin{scope}
\path[clip] (  0.00,  0.00) rectangle (433.62,180.67);

\path[] ( 44.94, 11.00) rectangle (410.11, 36.45);
\end{scope}
\begin{scope}
\path[clip] (  0.00,  0.00) rectangle (433.62,180.67);
\definecolor{drawColor}{RGB}{0,0,0}

\node[text=drawColor,anchor=base west,inner sep=0pt, outer sep=0pt, scale=  0.80] at ( 50.44, 20.97) {EMCB cluster (with $95\%$ confidence interval):};
\end{scope}
\begin{scope}
\path[clip] (  0.00,  0.00) rectangle (433.62,180.67);
\definecolor{drawColor}{RGB}{228,26,28}

\path[draw=drawColor,line width= 0.8pt,line join=round] (217.50, 23.73) -- (229.06, 23.73);
\end{scope}
\begin{scope}
\path[clip] (  0.00,  0.00) rectangle (433.62,180.67);
\definecolor{drawColor}{RGB}{228,26,28}
\definecolor{fillColor}{RGB}{51,51,51}

\path[draw=drawColor,line width= 0.6pt,fill=fillColor,fill opacity=0.15] (216.76, 17.21) rectangle (229.80, 30.24);
\end{scope}
\begin{scope}
\path[clip] (  0.00,  0.00) rectangle (433.62,180.67);
\definecolor{drawColor}{RGB}{55,126,184}

\path[draw=drawColor,line width= 0.8pt,line join=round] (277.61, 23.73) -- (289.17, 23.73);
\end{scope}
\begin{scope}
\path[clip] (  0.00,  0.00) rectangle (433.62,180.67);
\definecolor{drawColor}{RGB}{55,126,184}
\definecolor{fillColor}{RGB}{51,51,51}

\path[draw=drawColor,line width= 0.6pt,fill=fillColor,fill opacity=0.15] (276.88, 17.21) rectangle (289.91, 30.24);
\end{scope}
\begin{scope}
\path[clip] (  0.00,  0.00) rectangle (433.62,180.67);
\definecolor{drawColor}{RGB}{77,175,74}

\path[draw=drawColor,line width= 0.8pt,line join=round] (351.83, 23.73) -- (363.39, 23.73);
\end{scope}
\begin{scope}
\path[clip] (  0.00,  0.00) rectangle (433.62,180.67);
\definecolor{drawColor}{RGB}{77,175,74}
\definecolor{fillColor}{RGB}{51,51,51}

\path[draw=drawColor,line width= 0.6pt,fill=fillColor,fill opacity=0.15] (351.09, 17.21) rectangle (364.13, 30.24);
\end{scope}
\begin{scope}
\path[clip] (  0.00,  0.00) rectangle (433.62,180.67);
\definecolor{drawColor}{gray}{0.30}

\node[text=drawColor,anchor=base west,inner sep=0pt, outer sep=0pt, scale=  0.80] at (234.51, 20.97) {Low (2.03)};
\end{scope}
\begin{scope}
\path[clip] (  0.00,  0.00) rectangle (433.62,180.67);
\definecolor{drawColor}{gray}{0.30}

\node[text=drawColor,anchor=base west,inner sep=0pt, outer sep=0pt, scale=  0.80] at (294.62, 20.97) {Medium (3.15)};
\end{scope}
\begin{scope}
\path[clip] (  0.00,  0.00) rectangle (433.62,180.67);
\definecolor{drawColor}{gray}{0.30}

\node[text=drawColor,anchor=base west,inner sep=0pt, outer sep=0pt, scale=  0.80] at (368.84, 20.97) {High (4.2)};
\end{scope}
\end{tikzpicture}
 
    \caption{
        Regression model predicting importance change.        
        {\normalfont 
The left graph shows a significant main effect of \emph{decision-making (recognition)} ($\beta = 0.23$, $p < 0.001$), the right graph of \emph{sense-making (ease)} ($\beta = 0.17$, $p < 0.001$).
Marginal effects analysis indicates a significant moderating effect of \emph{EMCB cluster} for \emph{decision-making (recognition)} on the linear contrast, $\beta = -0.28$, $p = 0.011$, indicating that problem recognition has a larger effect on importance change for participants with lower EMCB scores.
        The quadratic contrast was not significant ($p = 0.79$). The interaction between \emph{EMCB cluster} and \emph{sense-making (ease)} was not significant for both the linear ($p = 0.46$) and quadratic contrasts ($p = 0.62$).
        The model explained 10.4\% of variance in \emph{Importance change} ($R^2 = 0.1039$, $F(8, 302) = 4.38$, $p < 0.001$).
    }}
    \label{fig:importancechange}
\end{figure*}

\paragraph{Recognition impacting decision making}
Recognizing gaps in both knowledge and decision-making considerations contributed to a change in importance evaluations across all EMCB groups.
The most frequent positive responses on \hyperref[q:influence]{Q34} came from the medium EMCB group with 85 ouf of 120 comments (70.83\%) indicating that recognizing a gap positively influenced decision-making. The high EMCB group followed with 72 ouf of 106 comments (67.92\%), while the low EMCB group had 44 out of 77 comments (57.14\%).
The following quotes illustrate how participants from these groups recognized such gaps and adjusted their views on ethical consumption:
\begin{center}
    \begin{tabular}{p{19pc}}
        \hline
        1 ``\textit{I didn't think it would matter to me, but now that I searched and read up about it, it makes me prefer to do business with that company more.}'' (DEI - Participant 27 - Low EMCB) \\
           2 ``\textit{I'll look into trying to make positive changes. For some reason I have the impression that 'ethically sourced' = overpriced and crap quality. Not a fair judgement to make when I've not done my research.}'' (Social Impact - Participant 234 - Low EMCB) \\
        3 ``\textit{Will be more aware to search for the source of the product and see how it is produced. It was a great eye opening seeing what happens even today. I wasn't aware it was that bad.}'' (Origin - Participant 68 - Medium EMCB) \\
        4 ``\textit{I knew it was important only in the back of my mind. I never considered actually including that into my research. This study changed how I will research products for the rest of my life, and I thank you for that!}'' (Labour - Participant 292 - High EMCB) \\
\hline
     \quad\quad\quad \quad\quad\quad\quad \quad{Recognition quotes - Positive change}
   \end{tabular}
\end{center}
These responses highlight how recognition prompted participants to rethink their purchase decisions. 
The first three quotes demonstrate how the realization of (previously) lacking knowledge corrects unawareness (quotes 1 and 3) and shifts preconceived notions (quote 2) through informed research.
The third and fourth quotes also reveal the lasting influence on decision-making considerations, showing how newfound ethical understanding transforms research behaviours and future priorities.
These responses emphasize how awareness of knowledge gaps on ethical considerations can catalyze changes in attitudes toward responsible consumption, including for participants in the low and medium EMCB groups.

\paragraph{Recognition not impacting decision-making}
A (newfound) understanding of ethical aspects for purchase decisions
did not consistently lead to changes in importance evaluations across the EMCB groups. 
The responses for \hyperref[q:influence]{Q34} were most frequent in the low EMCB group, where 29 out of 77 comments (37.66\%) indicated no effect on decision-making. In the medium EMCB group, 30 out of 120 comments (25.00\%) reflected this lack of influence, while the high EMCB group had the fewest such comments, with 11 out of 106 responses (10.38\%). The following quotes highlight various reasons why recognition did not affect participants' ethical considerations:

\begin{center}
    \begin{tabular}{p{19pc}}
        \hline
        5\label{q:rec_not_one} ``\textit{It hasn't. I already consider it when possible.}'' (DEI, Participant 17, High EMCB) \\
        6\label{q:rec_not_two} `\textit{It poorly influenced my decision making, as the details about ethical topics were not easy to access to.}'' (Sourcing, Participant 322, High EMCB) \\ 
        7\label{q:rec_not_three} ``\textit{It hasn't had much influence as a lot of what brands write is not true, therefore I am more concerned about the environmental impact, product quality, and price.}'' (Social Impact - Participant 154, Medium EMCB) \\
        8\label{q:rec_not_four} ``\textit{Not at all. I rarely ever consider child labour or employment rights when I buy things. Most products are manufactured in China so it is pretty much a given that forced labour has at one point happened.}'' (Labour, - Participant 190, Low EMCB) \\
\hline
        \quad\quad\quad \quad\quad\quad\quad \quad{Recognition quotes - Negative Change}
    \end{tabular}
\end{center}
These quotes reveal diverse barriers that prevented participants from re-evaluating ethical aspects in their decision-making.
Quote \ref{q:rec_not_one} shows that, although they acknowledged the aspect considered, their ethical considerations had already reached a plateau.
This aligns with the lack of effect of recognizing a gap for high-EMCB participants in Figure \ref{fig:importancechange}.
In quote \ref{q:rec_not_two} and \ref{q:rec_not_three},
a lack of accessible or trustworthy information was a significant barrier, preventing participants from integrating ethical aspects into their decision-making process and instead focusing on more easily accessible aspects.
Finally, quote \ref{q:rec_not_four} expressed a sense of inevitability, feeling that unethical practices were too systemic and widespread to be influenced by individual consumer choices.
These themes suggest that barriers such as lack of trustworthy information, perceived ineffectiveness, accessibility issues, and competing priorities can prevent 
recognised ethical gaps in knowledge and decision-making
from influencing consumer behaviour.

\paragraph{Ease of Sense-Making Leading to Importance Change}
In total 117 out of 316 comments on \hyperref[q:learning]{Q32} talked about sense making in relation to their decision making and the importance they placed on ethical considerations:

\begin{center}
    \begin{tabular}{p{19pc}}
        \hline
9\label{q:ease_one} ``\textit{As I considered and searched for eco-friendly sustainable practices when buying a Nike shoe, I learned some amazing things! [...] I was impressed by Nike's commitment to sustainability and transparency, making it easier for me to make an informed purchase.}'' (EcoFriendly – Participant 187 - High EMCB) \\    
10\label{q:ease_two} ``\textit{I learnt that with a small amount of clicks, one can 'investigate' almost every brand. Not all brands are 'saints' and if someone thinks that considering a brand's political stance and ideology is important before making a purchase, searching the net is an easy way to figure that out.}'' (Ideology – Participant 184 - Medium EMCB) \\
 11\label{q:ease_three} `\textit{Selecting the right product that suits low carbon foot print as indicated by the product manufacturer is now easy to identify through the search engine.}'' (EcoFriendly – Participant 100 - High EMCB) \\ 
12\label{q:ease_four} ``\textit{Firstly I was pleased and surprised that the information was out there and easier to find than I thought it would be. [...] I tend to think more in terms of food and clothes products regards ethical sourcing and production than I do electronics...}'' (Sourcing – Participant 15 - Medium EMCB) \\
        \hline
       \quad\quad\quad \quad{Ease of Sense-Making quotes - Positive Change}
    \end{tabular}
\end{center}

These four quotes highlight how participants used easily accessible sources, like company websites and search engines, to make sense of ethical information and incorporate it into their decisions.
Quote \ref{q:ease_one} shows how transparent sustainability practices enabled informed, responsible choices.
Quote \ref{q:ease_two} and \ref{q:ease_three} illustrate how readily available information about brand values and product sustainability empowered participants to confidently consider ethical aspects in their decisions.
Quote \ref{q:ease_four} reflects how access to information expanded awareness of ethical sourcing in less considered categories, like electronics.
Together, these quotes emphasize how ease of sense-making through accessible, trustworthy sources encourages participants to prioritize sustainability in their purchasing decisions.

\paragraph{Challenges in Sense-Making}
While some participants were able to integrate ethical considerations into their decision-making, many reported (185 out of 315 comments on \hyperref[q:challenges]{Q34}, 58.73\%) barriers in accessing and understanding the relevant information:
\begin{center}
    \begin{tabular}{p{19pc}}
        \hline
        13\label{q:ease_not_one} ``\textit{Companies may not directly disclose if their product is from an inclusive/diverse/equal workforce. You may need to do a lot of research to find any relevant informaiton about the product you purchase and the workforce you're interested in supporting.''} (DEI – Participant 325 - High EMCB) \\ 
         15\label{q:ease_not_two} ``\textit{Due to the nature of the information needed , some information are not straightforward as they appear to be a kind of marketing or advertisement. This posed a challenge in getting the right information to aid my decision.}'' (Sourcing – Participant 332 - High EMCB) \\ 
15\label{q:ease_not_three} ``\textit{It's hard to find relevant information and to find the right search terms and how to compare the results and if you can trust them.}'' (Political Stance / Ideology – Participant 66 - Low EMCB) \\ 
16\label{q:ease_not_four} ``\textit{That I'm less willing to search all information and cross source it because I just don't want to take that time after work, etc ... Those information should be easy to find and we should be able to trust those information.}'' (Origin – Participant 59 - High EMCB) \\ \hline
      \quad\quad\quad \quad {Difficulties in Sense Making quotes - Negative Change}
    \end{tabular}
\end{center}

These quotes reveal various difficulties in accessing and trusting information:
A lack of clear information, as highlighted in quote \ref{q:ease_not_one} and \ref{q:ease_not_two}, directly prevented participants from factoring in ethical aspects.
Challenges in finding the right search terms, as seen in quote \ref{q:ease_not_three}, show how unfamiliarity with terminology can block access to relevant information.
Time and effort constraints are emphasized by quotes \ref{q:ease_not_one} and \ref{q:ease_not_four}, where the difficulty of cross-checking information discouraged the consideration of ethical aspects.
Trust issues, evident in quote \ref{q:ease_not_two}--\ref{q:ease_not_four}, illustrate how doubts about reliability can further prevent participants from using relevant information.
Overall, these comments underscore how difficulties in sense-making -- whether due to inaccessible information, lack of transparency, or time constraints -- can prevent consumers from integrating ethical considerations into their purchasing decisions.

 \section{Discussion}\label{sec_discussion}
The present task-based study investigated the role of search in changing participants' valuation of ethical, social, and environmental (short: ethical) aspects during purchase decisions. It combined both quantitative and qualitative data to explore the complexities of how participants engage with ethical decision-making.
Even though an overall positive influence of the search task on importance change was observed (\ref{rq1}), this was not attributable to any particular product category nor any particular aspect(s) considered.
Moreover, it was not linked to participants' ethical intentions (\ref{rq2}a) or differences in search activity (\ref{rq2}b).
Instead, the key factors identified were participants' recognition of ethical gaps in their knowledge and decision making and their ability to make sense of the information and decision space (\ref{rq2}c).
These results support the idea of an intention-`search'-behaviour gap~\cite{azzopardi2024src} where ethical intentions do not directly translate into information search, and search does not necessarily lead to changes in importance valuations for these aspects during decision making.
Ultimately, it is the recognition and sense-making of ethical considerations, not ethical intentions or searching activity, that promotes shifts towards more responsible purchase decisions.

This study’s approach differs from earlier work in the field. 
Similar investigations often rely on surveys and observations, asking participants to retrospectively explain changes in observed behaviours. 
For example, in longitudinal surveys assessing shifts in consumer attitudes typically use retrospective self-reports~\cite{GregorySmith2013EmotionalDissonance,Hasanzade2018SelectingShopping, Casais2022ThePriorities,azzopardi2024src}. 
Alternatively, controlled lab studies test the effects of specific interventions in ceteris paribus settings, where all other variables are held constant. 
Such studies often isolate the influence of 
nudges or primes -- such as product attribute framing or ethical information availability -- 
on consumer choices~\cite{Green2014GreenShade, GregorySmith2013EmotionalDissonance}.
In contrast, our interactive, task-based study provided an ecologically valid setting~\cite{borlund2016}, capturing the broader complexity of searching for information and incorporating it into decision-making. 
By involving multiple, interrelated variables -- aspects, products, participants, queries, and documents -- over extensive search sessions, only robust effects were likely to surface. 
This approach, which included an online view of search behaviour and subjective reports of the search process, allowed us to identify the key search-related contributors to changes in responsible consumption attitudes.

Recognition of gaps in decision making emerged as a key driver of changes in the importance participants placed on ethical aspects. 
Surprisingly, this effect was more pronounced for participants with lower ethical intentions (low EMCB scores), who showed the greatest potential for change. 
Qualitative responses reflected this dynamic, as mostly low and medium EMCB participants described how recognizing gaps in their knowledge and decision-making process -- triggered by their searches -- led them to re-evaluate their priorities.
In contrast, participants with higher EMCB scores initially assigned more importance to ethical considerations.
Behavioural data showed that these participants engaged in more focused and extended searches, but that these efforts did not in themselves lead to significant changes in their purchase considerations.
This distinction aligns with earlier research on the intention-behaviour gap~\cite{carrigan2004better_shopping,Uusitalo2004EthicalFinland,Young2010SustainableProducts}, where stated intentions do not always lead to ethical behaviours. 
Instead, our findings suggest that not search effort per se, but problem recognition prompts participants to 
integrate ethical considerations into their decisions.

Sense-making emerged as second key driver of changes in the importance participants placed on ethical aspects.
When information was easy to find and evaluate, participants were more likely to adjust their priorities.
Interestingly, this impact was regardless of their prior ethical intentions.
Qualitative comments revealed that clear and transparent information about company practices empowered participants to make informed decisions.
Easy access to relevant details expanded participants’ awareness and helped them better integrate ethical considerations into their choices.
Conversely, when information was difficult to locate or unclear, participants struggled to prioritize ethical aspects, citing time and effort constraints as barriers to sense-making.
Behavioural data supports this, showing that smoother interaction -- fewer queries with more and longer engagement -- was linked to increased sense-making.
These findings align with earlier research on the challenges of information overload and lack of transparency in evaluating ethical claims~\cite{azzopardi2024src,Keller1987EffectsEffectiveness,Schleenbecker2015InformationCoffee,Wiederhold2018EthicalIndustry}.
They extend this research by showing that accessible, clear information directly increases the likelihood of participants incorporating ethical considerations into their decisions.

One limitation of this study is that the two subjective factors explain only 10\% of the variance in importance change. While modest, this is notable given the complexity of variables influencing how participants evaluate ethical aspects. Such levels of explained variance are likewise common in behavioural studies of information search~\citep{robertson2023,vandersluis2013combined}.
Another limitation is the potential for demand characteristics and socially desirable responses, 
as participants may provide answers they believe are expected rather than their true feelings. Although this effect appeared limited -- some participants openly stated they ``didn't care'' about certain aspects (see Supplement \hyperref[app:apathy]{G}) -- social desirability bias is well-documented in attitude research. Employing controlled experiments and implicit attitude tests \citep{baranan2014,albarracin2018} could therefore complement our findings.
Finally, our study monetarily incentivized participants to engage with ethical information. Naturally, such incentives are not feasible in everyday practice, which instead would need to rely on search results, prompts, or product metadata to expose users to ethical information.
These limitations highlight the need for additional research employing a variety of research designs ~\citep{vandersluis2017las} and interventions to understand how search can affect responsible decision-making.

Our findings show that for specific purchase decisions, the importance assigned to ethical aspects does not simply shift due to ethical intentions or subsequent search activity. Instead, it changes when participants recognize a decision problem and sense they can resolve it.
This finding aligns well recent findings \citep{vandersluis2025ipm} and sense-making theory \citep{savolainen1993sensemaking}, which posits that successful information seeking hinges on identifying and resolving a ``\textit{knowledge gap}'' of one's understanding of the information and decision space.
Contrarily, it challenges the conventional focus on intentions by highlighting the importance of concrete decision-making contexts. 
Accordingly, we propose viewing ethical consumption as a \textit{partial information problem}: when consumers can recognize and make sense of gaps in their decision-making, they are more likely to prioritize ethical considerations. This highlights the role of information, rather than intentions alone, in driving behavioural change.
This shifts attention away from the intention-behaviour gap to the specific decision-making and sense-making challenges that responsible consumers face. However, this \textit{partial} hypothesis acknowledges that factors such as price, quality, and consumers' purchasing power typically dominate and constrain consumer decision-making~\citep{azzopardi2024src}.
Following this hypothesis, responsible consumption becomes a practical problem, where providing consumers with the right information at the right moment is key to fostering more responsible decisions.

IR systems already play a critical role in helping users navigate complex purchasing decisions~\citep{Ghose2014ExaminingRevenue}. 
Our findings suggest that this role can expand to support responsible consumption as well by enhancing both decision-making and sense-making processes.
From an IR perspective, a major challenge is improving the retrieval of ethical information, often obscured by green-washing and information asymmetries. 
Retrieval techniques such as for fairness and viewpoint diversity~\citep{draws2023viewpoint,fang2024fairness} can help surface ethical information, transcending predominant commercial views.
Refining search relevance criteria to include ethical factors~\citep{sundin2021relevance} could further prioritize ethical considerations in search results.
These techniques can enhance exposure to ethical content, increasing chances for consumers to recognize and make sense of ethical concerns during product searches, providing accuracy and the right level of complexity~\citep{vandersluis2013combined,vandersluis2024jasist}.

From an interactive perspective, 
decision-making support can offer consumers a clearer overview of the decision space. 
Integrating ethical insights into product metadata, such as through sustainability tags \citep{jager2022greendb, Grebitus2020SustainableChoices}, knowledge context \citep{smith2019} or query priming \citep{yamamoto2018query_priming}, can make ethical aspects more prominent and accessible. 
Such interfaces can help users identify gaps in their knowledge by visualizing the ethical dimensions of their choices, while minimizing cognitive load to simplify complex trade-offs between sustainability, price, and quality~\citep{vandersluis2010}.
By integrating both retrieval enhancements and interactive supports that emphasize ethical aspects, search systems can equip consumers with the necessary knowledge and context to make informed purchasing decisions.
Future research can explore the effectiveness of these technologies to support ethical consumerism without adding undue burden.

 \section{Summary and Conclusion}\label{sec_conclusion}

Our findings offer a nuanced perspective on when search changes minds. 
Rather than being driven merely by search activity or ethical intentions, 
we find that search can positively affect consumer behaviour by raising awareness of decision-making gaps and helping users better understand their choices. 
This insight shifts the focus from the widely discussed intention-behaviour gap in the literature, proposing instead that responsible consumption should be viewed as a partial information problem. 
While it is 'partial' because pricing, availability, and economic constraints heavily influence consumer decisions, 
it is fundamentally an 'information problem':
Better information access can help reveal and bridge gaps in knowledge and awareness, allowing consumers to make more informed, aligned, and ideally more ethical, choices.

Solving the partial information problem necessitates systemic changes: 
producers and retailers need to highlight ethical factors, IR systems and platforms must facilitate search and comparison based on these criteria, and watchdogs and regulators must ensure transparency, enforce standards, and hold companies accountable for misleading claims or unethical practices. 
This includes promoting the availability of verified ethical information and certifications, so consumers can trust the data they receive.
Ultimately, collaboration across industries, platforms, and regulatory bodies is required to create a marketplace where responsible consumption becomes the norm.

The true potential of IR systems lies in fostering a more informed consumer base. As consumers gain more convenient access to clear, relevant information, their demand for ethical products will increase, pressuring businesses to adopt more sustainable practices. By supporting informed decisions, responsible IR systems can contribute to a fairer and transparent marketplace, that facilitates more informed choices and aligned purchasing decisions.

\bibliographystyle{ACM-Reference-Format}
\bibliography{references.bib}

\clearpage
\setcounter{page}{1}
\appendix
\setcounter{table}{0}
\renewcommand{\thetable}{S\arabic{table}}
\setcounter{figure}{1}
\renewcommand{\thefigure}{S\arabic{figure}}
\renewcommand{\thesection}{\Alph{section}}
\renewcommand{\thesubsection}{\alph{subsection})}
\setcounter{section}{0}

\section*{Supplementary Materials}

This document provides supplementary materials for \textbf{Search Changes Consumers' Minds: How Recognizing Gaps Drives Sustainable Choices} by Frans van der Sluis and Leif Azzopardi, presented at ACM CHIIR 2025.

These materials include additional details relevant to the study, structured as follows:
    \begin{enumerate}[label=\Alph*]
    	\item Survey questions
    	\item Ethical aspects
    	\item ChatGPT prompts
    	\item Extended model of importance change
        \item Participant clustering
        \item Variable clustering
        \item Apathy toward ethical considerations
    \end{enumerate}

\section{Ethical aspects}\label{app:aspects}
Table \ref{tab:aspects_overview} gives an overview of the aspects, their descriptions, and examples as used in the survey.

\begin{table*}[htb]
\centering
\begin{tabular}{p{.15\textwidth}p{.40\textwidth}p{.40\textwidth}}
\toprule
\textbf{Aspect} & \textbf{Description} & \textbf{Examples} \\
\midrule
Political Stance / Ideology & The messaging and stance of the company and what it promotes/pushes/believes in. & What political or social position/stance does the company have? Do they promote or push a particular agenda? \\
Ethical Sourcing / Production & How the product was made, and whether the components/materials were sourced ethically, suppliers paid a fair price, etc. & How was the product made? Were the components and materials sourced ethically, from responsible suppliers, who were paid fairly? \\
Slave/Child Labour / Employment Rights & The treatment and rights of workers and whether they are being exploited, the benefits they are provided, etc. & How well are workers treated? Are they forced or coerced into working? Are they paid fairly? What conditions do they work in? \\
Country of Origin / Place of Manufacture & The place/country where the products were made/produced. & Which factory, region, and country is the product made? Are these places known for high standards, good environmental practices, etc.? \\
Social Impact / Investing in Community & The charity, outreach, and support given to communities and organizations and what they are supporting. & What organizations and communities does the company support? What outreach and charity do they provide? \\
Eco Friendly / Sustainable Practices & The impact that the company has on the environment and ecology (e.g., sustainable growth, carbon neutral, etc.). & How environmentally friendly is the company? Do their factories pollute the water, use green energy, and what impact do they have on local ecology? \\
Governance \& Compliance & The legitimacy and lawfulness of the company, and whether it follows rules, legislation, standards, etc., and can be held accountable. & Is the company registered? Do they follow legal requirements? Are their accounts available? Do they reside in a tax haven? \\
Diversity / Equality / Inclusivity & The programs and support for diversity, equality, and inclusivity. & Do they have DEI programs? Are they equal opportunity employers? Do they have specific programs that favor one group over another? \\
\bottomrule
\end{tabular}
\caption{\label{tab:aspects_overview} Overview of aspects, their descriptions, and examples as shown to participants. }
\end{table*}

\section{Survey questions}\label{app:questions}

\newlist{questions}{enumerate}{1}
\setlist*[questions,1]{%
  label=Q\arabic*,
}

\newlist{answers}{enumerate*}{1}
\setlist*[answers,1]{%
  label=\alph*),
}

\paragraph{Part 1: Recent Purchase Questions} \label{app:questions1}
\begin{questions}

  \item \label{q:confirmation} \textbf{Confirmation}
    I confirm that I am in the process of making a purchase of at least £100/€100, but have not yet purchased the product.
    \begin{answers}
      \item Yes, I am eligible to participate in this study.
      \item No, I am not eligible to participate in this study.
    \end{answers}

  \item \label{q:productdesc} \textbf{Product Description}
    What is the product that you are currently considering purchasing? Provide a high-level description (e.g., phone, suit, BBQ, couch, etc.).

  \item \label{q:productcategory} \textbf{Product Category}
    What category is the product in?
    \begin{answers}
      \item Clothing, Shoes, Fashion Accessories
      \item Household appliances \& goods, furniture, etc.
      \item Consumer electronics
      \item Books, movies, games, toys
      \item Sports, Recreation, Hobbies, etc.
      \item DIY \& Garden
      \item Other
    \end{answers}

  \item \label{q:decisionstage} \textbf{Purchasing Decision Stage}
    Where are you in terms of your purchasing decision?
    \begin{answers}
      \item Starting to look and browse for products that meet my needs
      \item Searching for different alternatives that meet my needs
      \item Comparing and researching specific alternatives
      \item Trying to decide which alternative to purchase
      \item Looking for the best possible deal for the product that I have selected to purchase
      \item I have purchased the product
    \end{answers}

  \item \label{q:timespent} \textbf{Time Spent Researching}
    How many hours in total have you spent researching the product and its alternatives so far?
    \begin{answers}
      \item Less than 1 hour
      \item 1-2 hours
      \item 2-4 hours
      \item 5-8 hours
      \item 9-24 hours
      \item 24 hours or more
    \end{answers}

  \item \label{q:researchperiod} \textbf{Research Period}
    Over what period have you spent researching the product and its alternatives?
    \begin{answers}
      \item Over the course of a day
      \item Over the course of a week
      \item Over the course of a month
      \item Longer than a month
    \end{answers}

  \item \label{q:alternatives} \textbf{Number of Alternatives Considered}
    How many alternatives (different products) have you considered so far?
    \begin{answers}
      \item 1
      \item 2-3
      \item 4-5
      \item 6-9
      \item 10 or more
    \end{answers}

\end{questions}

The following questions pertain to a specific aspect that you are considering while shopping:

\begin{questions}[resume]

  \item \label{q:importance} \textbf{Importance}
    How important is "[Aspect Name]" to you when comparing products and their alternatives?
    \begin{answers}
      \item Not at all important
      \item Slightly important
      \item Moderately important
      \item Very important
      \item Extremely important
    \end{answers}

  \item \label{q:considered} \textbf{Consideration}
    Have you considered issues associated with "[Aspect Name]" when searching and evaluating products?
    \begin{answers}
      \item Yes, I have been considering it
      \item No, I have not been considering it
    \end{answers}

  \item \label{q:searched} \textbf{Searched for Information}
    Have you specifically looked for information about "[Aspect Name]" when comparing products and brands?
    \begin{answers}
      \item Yes, I have (e.g., using a search engine to find reviews, ratings, certifications, etc.)
      \item No, I have not
    \end{answers}

\end{questions}

\paragraph{Part 2: Search Task}\label{app:questions2}
The following are the search instructions provided to participants:
\begin{questions}[resume]

  \item \label{q:searchtask} \textbf{Search Task}
    Using our search engine, we would like you to search for information regarding "[Aspect Name]" given the products (e.g., "[Product Name]") that you are shopping for.

    That is, we would like you to find out what each product/company offers, provides, or promotes with respect to "[Aspect Name]," which considers "[Aspect Description]."

    For example, you might want to find out about "[Aspect Examples]."

    If you don’t know much about this aspect, try searching for information to learn more about "[Aspect Examples]," and then search for product-specific information.

    \item \label{q:searchtaskdetails} \textbf{Search Instructions}
    \begin{enumerate}
      \item Search for approximately 10 minutes to compare the different products/companies you have been considering.
      \item Try to find at least 5 websites or pages that contain relevant information.
    \end{enumerate}
    
    \textbf{Important:} Make sure that you search via the provided link. If you do not perform the search task, your survey answers will not be eligible.

  \item \label{q:searchcompleted} \textbf{Have you completed the search task?}
    \begin{answers}
      \item Yes, I have completed the search task.
      \item No, I did not complete the search task.
    \end{answers}

\end{questions}

The following questions were designed to capture participants’ responses in each phase of the consumer purchase process.
\paragraph{Problem Recognition}

\begin{questions}[resume]
  \item \label{q:keydecision} \textbf{Key to Decision}
    I wanted to know about "[Aspect Name]" as it was key to my decision making.
    \begin{answers}
      \item Strongly disagree
      \item Disagree
      \item Somewhat disagree
      \item Neither agree nor disagree
      \item Somewhat agree
      \item Agree
      \item Strongly agree
    \end{answers}

  \item \label{q:knowledgeissues} \textbf{Knowledge Issues}
    I knew too little about "[Aspect Name]" to make my purchasing decision.
    \begin{answers}
      \item Strongly disagree
      \item Disagree
      \item Somewhat disagree
      \item Neither agree nor disagree
      \item Somewhat agree
      \item Agree
      \item Strongly agree
    \end{answers}

  \item \label{q:moresearching} \textbf{More Searching}
    I need to perform more searching about "[Aspect Name]" to inform my decision making.
    \begin{answers}
      \item Strongly disagree
      \item Disagree
      \item Somewhat disagree
      \item Neither agree nor disagree
      \item Somewhat agree
      \item Agree
      \item Strongly agree
    \end{answers}

  \item \label{q:moreresearchneeded} \textbf{More Research Needed}
    I need to perform more research into "[Aspect Name]" to make a good decision.
    \begin{answers}
      \item Strongly disagree
      \item Disagree
      \item Somewhat disagree
      \item Neither agree nor disagree
      \item Somewhat agree
      \item Agree
      \item Strongly agree
    \end{answers}
\end{questions}

\paragraph{Information Search}

\begin{questions}[resume]
  \item \label{q:easyfind} \textbf{Easy to Find Information}
    I found it easy to find any information about "[Aspect Name]".
    \begin{answers}
      \item Strongly disagree
      \item Disagree
      \item Somewhat disagree
      \item Neither agree nor disagree
      \item Somewhat agree
      \item Agree
      \item Strongly agree
    \end{answers}

  \item \label{q:lotsofinfo} \textbf{Lots of Information}
    I could see that there was a lot of relevant information about "[Aspect Name]" available online.
    \begin{answers}
      \item Strongly disagree
      \item Disagree
      \item Somewhat disagree
      \item Neither agree nor disagree
      \item Somewhat agree
      \item Agree
      \item Strongly agree
    \end{answers}

  \item \label{q:searchsuccess} \textbf{Search Success}
    I was not successful in finding any relevant information about "[Aspect Name]".
    \begin{answers}
      \item Strongly disagree
      \item Disagree
      \item Somewhat disagree
      \item Neither agree nor disagree
      \item Somewhat agree
      \item Agree
      \item Strongly agree
    \end{answers}

  \item \label{q:timeconsuming} \textbf{Time Consuming}
    I think it would be very time-consuming to find all the relevant information about "[Aspect Name]".
    \begin{answers}
      \item Strongly disagree
      \item Disagree
      \item Somewhat disagree
      \item Neither agree nor disagree
      \item Somewhat agree
      \item Agree
      \item Strongly agree
    \end{answers}
\end{questions}

\paragraph{Information Evaluation}

\begin{questions}[resume]
  \item \label{q:sensemaking} \textbf{Sense-making Issues}
    I could not make sense of the information I found about "[Aspect Name]".
    \begin{answers}
      \item Strongly disagree
      \item Disagree
      \item Somewhat disagree
      \item Neither agree nor disagree
      \item Somewhat agree
      \item Agree
      \item Strongly agree
    \end{answers}

  \item \label{q:trustinfo} \textbf{Trust in Information}
    I could trust the information I found about "[Aspect Name]".
    \begin{answers}
      \item Strongly disagree
      \item Disagree
      \item Somewhat disagree
      \item Neither agree nor disagree
      \item Somewhat agree
      \item Agree
      \item Strongly agree
    \end{answers}

  \item \label{q:completepicture} \textbf{Complete Picture}
    I was able to build up a complete picture of "[Aspect Name]" for my purchasing decision.
    \begin{answers}
      \item Strongly disagree
      \item Disagree
      \item Somewhat disagree
      \item Neither agree nor disagree
      \item Somewhat agree
      \item Agree
      \item Strongly agree
    \end{answers}

  \item \label{q:toomuchinfo} \textbf{Too Much Information}
    I felt there was too much information about "[Aspect Name]" to check, consider and compare.
    \begin{answers}
      \item Strongly disagree
      \item Disagree
      \item Somewhat disagree
      \item Neither agree nor disagree
      \item Somewhat agree
      \item Agree
      \item Strongly agree
    \end{answers}
\end{questions}

\paragraph{Decision-Making}

\begin{questions}[resume]
  \item \label{q:wontuse} \textbf{Won't Use Information}
    I won’t be using any of the information about "[Aspect Name]" to make my purchasing decision.
    \begin{answers}
      \item Strongly disagree
      \item Disagree
      \item Somewhat disagree
      \item Neither agree nor disagree
      \item Somewhat agree
      \item Agree
      \item Strongly agree
    \end{answers}

  \item \label{q:hardcompare} \textbf{Hard to Compare Products}
    I found it hard to compare products using information about "[Aspect Name]".
    \begin{answers}
      \item Strongly disagree
      \item Disagree
      \item Somewhat disagree
      \item Neither agree nor disagree
      \item Somewhat agree
      \item Agree
      \item Strongly agree
    \end{answers}

  \item \label{q:uncertainty} \textbf{Uncertainty about Decision}
    I am uncertain about which product to buy given the information I found about "[Aspect Name]".
    \begin{answers}
      \item Strongly disagree
      \item Disagree
      \item Somewhat disagree
      \item Neither agree nor disagree
      \item Somewhat agree
      \item Agree
      \item Strongly agree
    \end{answers}

  \item \label{q:difficulttradeoffs} \textbf{Difficult Trade-offs}
    I will have to make some difficult trade-offs given information about "[Aspect Name]".
    \begin{answers}
      \item Strongly disagree
      \item Disagree
      \item Somewhat disagree
      \item Neither agree nor disagree
      \item Somewhat agree
      \item Agree
      \item Strongly agree
    \end{answers}

  \item \label{q:reconsidering} \textbf{Reconsidering}
    I am reconsidering which product to buy after finding out about "[Aspect Name]".
    \begin{answers}
      \item Strongly disagree
      \item Disagree
      \item Somewhat disagree
      \item Neither agree nor disagree
      \item Somewhat agree
      \item Agree
      \item Strongly agree
    \end{answers}
\end{questions}

\paragraph{Part 3: Search Impact and EMCB Questionnaire} \label{app:questions3}
\begin{questions}[resume]

  \item \label{q:updatedimportance} \textbf{Updated Importance}
    After considering and searching for information about "[Aspect Name]", how important do you now think it is when comparing products and their alternatives?
    \begin{answers}
      \item Not at all important
      \item Slightly important
      \item Moderately important
      \item Very important
      \item Extremely important
    \end{answers}

  \item \label{q:learning} \textbf{Learning}
    What did you learn from considering and searching about "[Aspect Name]"?

  \item \label{q:influence} \textbf{Influence on Decision}
    How has considering and searching for "[Aspect Name]" influenced your decision-making?

  \item \label{q:challenges} \textbf{Challenges}
    When searching for information online when shopping and comparing products, what are the biggest problems or challenges that you face?

\end{questions}

After these questions, participants answered the EMCB questionnaire \citep{SUDBURYRILEY20162697}.

\section{ChatGPT prompts}\label{app:chatgpt}
The prompts provided to ChatGPT to assist in extracting and summarizing qualitative data were as follows:

\begin{center}
\begin{tabular}{p{.85\linewidth}}
\midrule
\textit{Given the list of comments on what people learned regarding ethical aspects to their purchasing decision, extract comments that describe how they could (or could not) make sense of the information available (or not available) for an aspect.}\\
\textit{Extract both positive comments, where (the ease of) sense-making contributed to a change in their valuation of the importance of sustainable/responsible aspects, as well as negative comments, where (difficulties in) sense-making hindered an impact on their valuation.}\\
\textit{Comments: ``[comments]''}\\
\textit{Instructions:}\\
\textit{- Provide up to five comments as examples for each category.}\\
\textit{- Select comments that are the most prototypical and memorable, or provide explanations.}\\
\textit{- Give an indication of how many comments fit under each category.} \\
\midrule
\multicolumn{1}{r}{Sense-Making as a Driver of Importance Change}
\end{tabular}
\end{center}

\begin{center}
\begin{tabular}{p{.85\linewidth}}
\midrule
\textit{Given the list of comments on whether people changed their mind regarding their purchasing decision, extract comments that describe people recognizing a gap in their understanding.}\\
\textit{Extract both positive comments, where a gap contributed to a change in their valuation of the importance of sustainable/responsible aspects, as well as negative comments, where recognition of a gap did not impact their valuation.}\\
\textit{Comments: ``[comments]''}\\
\textit{Instructions:}\\
\textit{- Provide up to five comments as examples for each category.}\\
\textit{- Select comments that are the most prototypical and memorable, or provide explanations.}\\
\textit{- Give an indication of how many comments fit under each category.} \\
\midrule
\multicolumn{1}{r}{Recognition as a Driver of Importance Change}
\end{tabular}
\end{center}

These two prompts were used in conjunction with OpenAI's gpt-4o model to extract examples from participants' comments. All examples were verified against our data to prevent inaccuracies.

\begin{center}
\begin{tabular}{p{.85\linewidth}}
\midrule
\textit{Given the comment below, assign the following tags that are relevant:}

\textit{1. Positive Comments (Ease of Sense Making): Comments that describe how people could make sense of the information available for an aspect, or where (the ease of) sense making contributed to a change in their valuation of the importance of sustainable/responsible/ethical aspects.}

\textit{2. Negative Comments (Difficulties in Sense Making): Comments that describe how people could not make sense of the information, or the information was not available for an aspect, or where (difficulties in) sense making hindered an impact on their valuation of the importance of sustainable/responsible/ethical aspects.}

\textit{3. Positive Comments (Decision-Making Recognition): Comments that describe people who changed their mind regarding their purchasing decision, recognized a gap in their understanding about an aspect, or where a gap contributed to a change in their valuation of the importance of sustainable/responsible aspects.}

\textit{4. Negative Comments (No Decision-Making Recognition): Comments that describe people who did not change their mind regarding their purchasing decision, did not recognize a gap in their understanding about an aspect, or where a gap did not contribute to a change in their valuation of the importance of sustainable/responsible aspects.}

\textit{5. Other Comments (Not about Sense Making or Decision Making): Other comments that do not fit into the previous categories.}

\textit{Comment: ``[comment]''}

\textit{Instructions:}\\
\textit{- List the most relevant tags that apply to the comment.}\\
\textit{- Only output the tag numbers and names (not descriptions).}\\
\textit{- Output in a comma-separated format.} \\
\midrule
\multicolumn{1}{r}{Comment Tagging and Categorization}
\end{tabular}
\end{center}

This prompt was used in conjunction with OpenAI's gpt-3.5-turbo model to assign codes (labels) to participants' comments.

\section{Extended model of importance change}\label{app:extended_model}

A possible, alternative explanation for the non-significant effects 
of EMCB on changes in importance in Figure \ref{fig:boxplots}b
is that the influence of EMCB is more pronounced for specific aspects, particularly those related to social and environmental concerns.
A combined ANOVA confirmed this possibility with a minor interaction between aspect and EMCB cluster, $F(14, 287) = 1.99$, $p = .018$, alongside the main effects of both aspect, $F(7, 287) = 3.35$, $p = .002$, and EMCB cluster, $F(2, 287) = 83.07$, $p < .001$ on importance ratings.
Despite these effects on importance, no significant effect was observed for importance change, as neither aspect, $F(7, 287) = 1.65$, $p = .122$, EMCB cluster, $F(2, 287) = 1.48$, $p = .229$, nor their interaction, $F(14, 287) = 1.21$, $p = .270$, showed a significant influence on the change in importance.
These results
failed to support \ref{rq2}a for specific aspects.
This leaves open the question of what drove the observed changes in importance valuations.
If these changes are not attributable to individual intentions, they may instead be influenced by the search process itself.

\section{Participant clustering}\label{app:participant_clustering}
\begin{figure}[htb]
        \centering
        \begin{tikzpicture}[x=1pt,y=1pt]
\definecolor{fillColor}{RGB}{255,255,255}
\path[use as bounding box,fill=fillColor,fill opacity=0.00] (0,0) rectangle (216.81,108.41);
\begin{scope}
\path[clip] (  0.00,  0.00) rectangle (216.81,108.41);

\path[] (  0.00,  0.00) rectangle (216.81,108.41);
\end{scope}
\begin{scope}
\path[clip] ( 36.26, 27.33) rectangle (211.31,102.90);

\path[] ( 36.26, 27.33) rectangle (211.31,102.91);
\definecolor{drawColor}{RGB}{211,211,211}

\path[draw=drawColor,line width= 0.6pt,line join=round] ( 36.26, 28.11) --
	(211.31, 28.11);

\path[draw=drawColor,line width= 0.6pt,line join=round] ( 36.26, 45.26) --
	(211.31, 45.26);

\path[draw=drawColor,line width= 0.6pt,line join=round] ( 36.26, 62.42) --
	(211.31, 62.42);

\path[draw=drawColor,line width= 0.6pt,line join=round] ( 36.26, 79.57) --
	(211.31, 79.57);

\path[draw=drawColor,line width= 0.6pt,line join=round] ( 36.26, 96.73) --
	(211.31, 96.73);

\path[draw=drawColor,line width= 0.6pt,line join=round] ( 44.22, 27.33) --
	( 44.22,102.90);

\path[draw=drawColor,line width= 0.6pt,line join=round] ( 61.90, 27.33) --
	( 61.90,102.90);

\path[draw=drawColor,line width= 0.6pt,line join=round] ( 79.58, 27.33) --
	( 79.58,102.90);

\path[draw=drawColor,line width= 0.6pt,line join=round] ( 97.26, 27.33) --
	( 97.26,102.90);

\path[draw=drawColor,line width= 0.6pt,line join=round] (114.94, 27.33) --
	(114.94,102.90);

\path[draw=drawColor,line width= 0.6pt,line join=round] (132.63, 27.33) --
	(132.63,102.90);

\path[draw=drawColor,line width= 0.6pt,line join=round] (150.31, 27.33) --
	(150.31,102.90);

\path[draw=drawColor,line width= 0.6pt,line join=round] (167.99, 27.33) --
	(167.99,102.90);

\path[draw=drawColor,line width= 0.6pt,line join=round] (185.67, 27.33) --
	(185.67,102.90);

\path[draw=drawColor,line width= 0.6pt,line join=round] (203.35, 27.33) --
	(203.35,102.90);
\definecolor{drawColor}{RGB}{55,126,184}

\path[draw=drawColor,line width= 0.6pt,dash pattern=on 4pt off 4pt ,line join=round] ( 44.22, 99.47) --
	( 61.90, 55.52) --
	( 79.58, 45.85) --
	( 97.26, 41.77) --
	(114.94, 38.85) --
	(132.63, 36.15) --
	(150.31, 35.18) --
	(167.99, 33.45) --
	(185.67, 32.56) --
	(203.35, 30.77);
\definecolor{drawColor}{RGB}{0,0,0}
\definecolor{fillColor}{RGB}{0,0,0}

\path[draw=drawColor,line width= 0.4pt,line join=round,line cap=round,fill=fillColor] ( 44.22, 99.47) circle (  1.43);

\path[draw=drawColor,line width= 0.4pt,line join=round,line cap=round,fill=fillColor] ( 61.90, 55.52) circle (  1.43);

\path[draw=drawColor,line width= 0.4pt,line join=round,line cap=round,fill=fillColor] ( 79.58, 45.85) circle (  1.43);

\path[draw=drawColor,line width= 0.4pt,line join=round,line cap=round,fill=fillColor] ( 97.26, 41.77) circle (  1.43);

\path[draw=drawColor,line width= 0.4pt,line join=round,line cap=round,fill=fillColor] (114.94, 38.85) circle (  1.43);

\path[draw=drawColor,line width= 0.4pt,line join=round,line cap=round,fill=fillColor] (132.63, 36.15) circle (  1.43);

\path[draw=drawColor,line width= 0.4pt,line join=round,line cap=round,fill=fillColor] (150.31, 35.18) circle (  1.43);

\path[draw=drawColor,line width= 0.4pt,line join=round,line cap=round,fill=fillColor] (167.99, 33.45) circle (  1.43);

\path[draw=drawColor,line width= 0.4pt,line join=round,line cap=round,fill=fillColor] (185.67, 32.56) circle (  1.43);

\path[draw=drawColor,line width= 0.4pt,line join=round,line cap=round,fill=fillColor] (203.35, 30.77) circle (  1.43);
\end{scope}
\begin{scope}
\path[clip] (  0.00,  0.00) rectangle (216.81,108.41);
\definecolor{drawColor}{RGB}{0,0,0}

\node[text=drawColor,anchor=base east,inner sep=0pt, outer sep=0pt, scale=  0.80] at ( 31.31, 25.35) {1000};

\node[text=drawColor,anchor=base east,inner sep=0pt, outer sep=0pt, scale=  0.80] at ( 31.31, 42.51) {1500};

\node[text=drawColor,anchor=base east,inner sep=0pt, outer sep=0pt, scale=  0.80] at ( 31.31, 59.66) {2000};

\node[text=drawColor,anchor=base east,inner sep=0pt, outer sep=0pt, scale=  0.80] at ( 31.31, 76.82) {2500};

\node[text=drawColor,anchor=base east,inner sep=0pt, outer sep=0pt, scale=  0.80] at ( 31.31, 93.97) {3000};
\end{scope}
\begin{scope}
\path[clip] (  0.00,  0.00) rectangle (216.81,108.41);
\definecolor{drawColor}{gray}{0.20}

\path[draw=drawColor,line width= 0.6pt,line join=round] ( 33.51, 28.11) --
	( 36.26, 28.11);

\path[draw=drawColor,line width= 0.6pt,line join=round] ( 33.51, 45.26) --
	( 36.26, 45.26);

\path[draw=drawColor,line width= 0.6pt,line join=round] ( 33.51, 62.42) --
	( 36.26, 62.42);

\path[draw=drawColor,line width= 0.6pt,line join=round] ( 33.51, 79.57) --
	( 36.26, 79.57);

\path[draw=drawColor,line width= 0.6pt,line join=round] ( 33.51, 96.73) --
	( 36.26, 96.73);
\end{scope}
\begin{scope}
\path[clip] (  0.00,  0.00) rectangle (216.81,108.41);
\definecolor{drawColor}{gray}{0.20}

\path[draw=drawColor,line width= 0.6pt,line join=round] ( 44.22, 24.58) --
	( 44.22, 27.33);

\path[draw=drawColor,line width= 0.6pt,line join=round] ( 61.90, 24.58) --
	( 61.90, 27.33);

\path[draw=drawColor,line width= 0.6pt,line join=round] ( 79.58, 24.58) --
	( 79.58, 27.33);

\path[draw=drawColor,line width= 0.6pt,line join=round] ( 97.26, 24.58) --
	( 97.26, 27.33);

\path[draw=drawColor,line width= 0.6pt,line join=round] (114.94, 24.58) --
	(114.94, 27.33);

\path[draw=drawColor,line width= 0.6pt,line join=round] (132.63, 24.58) --
	(132.63, 27.33);

\path[draw=drawColor,line width= 0.6pt,line join=round] (150.31, 24.58) --
	(150.31, 27.33);

\path[draw=drawColor,line width= 0.6pt,line join=round] (167.99, 24.58) --
	(167.99, 27.33);

\path[draw=drawColor,line width= 0.6pt,line join=round] (185.67, 24.58) --
	(185.67, 27.33);

\path[draw=drawColor,line width= 0.6pt,line join=round] (203.35, 24.58) --
	(203.35, 27.33);
\end{scope}
\begin{scope}
\path[clip] (  0.00,  0.00) rectangle (216.81,108.41);
\definecolor{drawColor}{RGB}{0,0,0}

\node[text=drawColor,anchor=base,inner sep=0pt, outer sep=0pt, scale=  0.80] at ( 44.22, 16.87) {1};

\node[text=drawColor,anchor=base,inner sep=0pt, outer sep=0pt, scale=  0.80] at ( 61.90, 16.87) {2};

\node[text=drawColor,anchor=base,inner sep=0pt, outer sep=0pt, scale=  0.80] at ( 79.58, 16.87) {3};

\node[text=drawColor,anchor=base,inner sep=0pt, outer sep=0pt, scale=  0.80] at ( 97.26, 16.87) {4};

\node[text=drawColor,anchor=base,inner sep=0pt, outer sep=0pt, scale=  0.80] at (114.94, 16.87) {5};

\node[text=drawColor,anchor=base,inner sep=0pt, outer sep=0pt, scale=  0.80] at (132.63, 16.87) {6};

\node[text=drawColor,anchor=base,inner sep=0pt, outer sep=0pt, scale=  0.80] at (150.31, 16.87) {7};

\node[text=drawColor,anchor=base,inner sep=0pt, outer sep=0pt, scale=  0.80] at (167.99, 16.87) {8};

\node[text=drawColor,anchor=base,inner sep=0pt, outer sep=0pt, scale=  0.80] at (185.67, 16.87) {9};

\node[text=drawColor,anchor=base,inner sep=0pt, outer sep=0pt, scale=  0.80] at (203.35, 16.87) {10};
\end{scope}
\begin{scope}
\path[clip] (  0.00,  0.00) rectangle (216.81,108.41);
\definecolor{drawColor}{RGB}{0,0,0}

\node[text=drawColor,anchor=base,inner sep=0pt, outer sep=0pt, scale=  0.80] at (123.79,  7.06) {Number of Clusters};
\end{scope}
\begin{scope}
\path[clip] (  0.00,  0.00) rectangle (216.81,108.41);
\definecolor{drawColor}{RGB}{0,0,0}

\node[text=drawColor,rotate= 90.00,anchor=base,inner sep=0pt, outer sep=0pt, scale=  0.80] at ( 11.01, 65.12) {Total Within-Cluster Sum of Squares (WSS)};
\end{scope}
\end{tikzpicture}

\caption{  
    Scree plots for clustering analysis of EMCB scores. The Within-Cluster Sum of Squares (WCSS) score is shown on the y-axis, and the number of clusters is shown on the x-axis.}  

\label{fig:pca_scree}  

\end{figure}
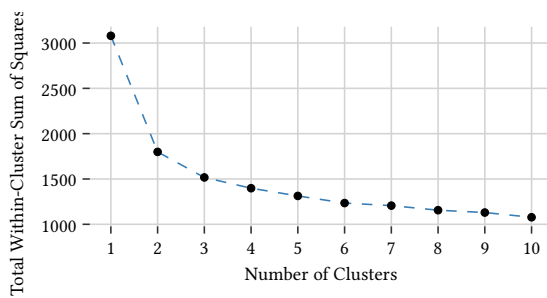

K-means clustering was applied to participants' EMCB scores to identify ethical consumption behaviour groupings. The elbow method suggested two or three clusters (see Figure \ref{fig:pca_scree}), while silhouette scores peaked at 0.34 for two clusters and 0.22 for three. Three clusters were retained, broadly corresponding to groupings of low, medium and high ethical mindedness, with strong separation and cohesion as shown in Figure \ref{fig:emcb}.
To visualize the clustering structure, Principal Component Analysis (PCA) was performed, reducing the EMCB scores to two principal components. This allowed for a 2-dimensional representation while preserving the variance in ethical intentions in Figure \ref{fig:emcb}.

\section{Variable clustering}\label{app:variable_clustering}
Exploratory Factor Analysis (EFA) with Promax rotation was conducted to identify latent constructs, enabling visualization of variable relationships in Figure \ref{fig:lfs_clustering}.
A two-factor solution was preferred for interpretability and visualization, supported by Kaiser's criterion (eigenvalues > 1). The first two factors had eigenvalues of 5.35 and 2.77, while the third dropped to 0.70, well below the retention threshold. 

\begin{figure}[htb]
    \centering
        \centering
        \begin{tikzpicture}[x=1pt,y=1pt]
\definecolor{fillColor}{RGB}{255,255,255}
\path[use as bounding box,fill=fillColor,fill opacity=0.00] (0,0) rectangle (216.81,108.41);
\begin{scope}
\path[clip] (  0.00,  0.00) rectangle (216.81,108.41);

\path[] (  0.00,  0.00) rectangle (216.81,108.41);
\end{scope}
\begin{scope}
\path[clip] ( 24.26, 27.33) rectangle (211.31,102.90);

\path[] ( 24.26, 27.33) rectangle (211.31,102.91);
\definecolor{drawColor}{RGB}{211,211,211}

\path[draw=drawColor,line width= 0.6pt,line join=round] ( 24.26, 30.30) --
	(211.31, 30.30);

\path[draw=drawColor,line width= 0.6pt,line join=round] ( 24.26, 54.79) --
	(211.31, 54.79);

\path[draw=drawColor,line width= 0.6pt,line join=round] ( 24.26, 79.29) --
	(211.31, 79.29);

\path[draw=drawColor,line width= 0.6pt,line join=round] ( 32.77, 27.33) --
	( 32.77,102.90);

\path[draw=drawColor,line width= 0.6pt,line join=round] ( 51.66, 27.33) --
	( 51.66,102.90);

\path[draw=drawColor,line width= 0.6pt,line join=round] ( 70.55, 27.33) --
	( 70.55,102.90);

\path[draw=drawColor,line width= 0.6pt,line join=round] ( 89.45, 27.33) --
	( 89.45,102.90);

\path[draw=drawColor,line width= 0.6pt,line join=round] (108.34, 27.33) --
	(108.34,102.90);

\path[draw=drawColor,line width= 0.6pt,line join=round] (127.23, 27.33) --
	(127.23,102.90);

\path[draw=drawColor,line width= 0.6pt,line join=round] (146.13, 27.33) --
	(146.13,102.90);

\path[draw=drawColor,line width= 0.6pt,line join=round] (165.02, 27.33) --
	(165.02,102.90);

\path[draw=drawColor,line width= 0.6pt,line join=round] (183.91, 27.33) --
	(183.91,102.90);

\path[draw=drawColor,line width= 0.6pt,line join=round] (202.81, 27.33) --
	(202.81,102.90);
\definecolor{drawColor}{RGB}{55,126,184}

\path[draw=drawColor,line width= 0.6pt,dash pattern=on 4pt off 4pt ,line join=round] ( 32.77, 99.47) --
	( 51.66, 45.82) --
	( 70.55, 39.06) --
	( 89.45, 35.63) --
	(108.34, 33.86) --
	(127.23, 32.49) --
	(146.13, 31.81) --
	(165.02, 31.27) --
	(183.91, 30.95) --
	(202.81, 30.77);
\definecolor{drawColor}{RGB}{0,0,0}
\definecolor{fillColor}{RGB}{0,0,0}

\path[draw=drawColor,line width= 0.4pt,line join=round,line cap=round,fill=fillColor] ( 32.77, 99.47) circle (  1.43);

\path[draw=drawColor,line width= 0.4pt,line join=round,line cap=round,fill=fillColor] ( 51.66, 45.82) circle (  1.43);

\path[draw=drawColor,line width= 0.4pt,line join=round,line cap=round,fill=fillColor] ( 70.55, 39.06) circle (  1.43);

\path[draw=drawColor,line width= 0.4pt,line join=round,line cap=round,fill=fillColor] ( 89.45, 35.63) circle (  1.43);

\path[draw=drawColor,line width= 0.4pt,line join=round,line cap=round,fill=fillColor] (108.34, 33.86) circle (  1.43);

\path[draw=drawColor,line width= 0.4pt,line join=round,line cap=round,fill=fillColor] (127.23, 32.49) circle (  1.43);

\path[draw=drawColor,line width= 0.4pt,line join=round,line cap=round,fill=fillColor] (146.13, 31.81) circle (  1.43);

\path[draw=drawColor,line width= 0.4pt,line join=round,line cap=round,fill=fillColor] (165.02, 31.27) circle (  1.43);

\path[draw=drawColor,line width= 0.4pt,line join=round,line cap=round,fill=fillColor] (183.91, 30.95) circle (  1.43);

\path[draw=drawColor,line width= 0.4pt,line join=round,line cap=round,fill=fillColor] (202.81, 30.77) circle (  1.43);
\end{scope}
\begin{scope}
\path[clip] (  0.00,  0.00) rectangle (216.81,108.41);
\definecolor{drawColor}{RGB}{0,0,0}

\node[text=drawColor,anchor=base east,inner sep=0pt, outer sep=0pt, scale=  0.80] at ( 19.31, 27.54) {0};

\node[text=drawColor,anchor=base east,inner sep=0pt, outer sep=0pt, scale=  0.80] at ( 19.31, 52.04) {1};

\node[text=drawColor,anchor=base east,inner sep=0pt, outer sep=0pt, scale=  0.80] at ( 19.31, 76.53) {2};
\end{scope}
\begin{scope}
\path[clip] (  0.00,  0.00) rectangle (216.81,108.41);
\definecolor{drawColor}{gray}{0.20}

\path[draw=drawColor,line width= 0.6pt,line join=round] ( 21.51, 30.30) --
	( 24.26, 30.30);

\path[draw=drawColor,line width= 0.6pt,line join=round] ( 21.51, 54.79) --
	( 24.26, 54.79);

\path[draw=drawColor,line width= 0.6pt,line join=round] ( 21.51, 79.29) --
	( 24.26, 79.29);
\end{scope}
\begin{scope}
\path[clip] (  0.00,  0.00) rectangle (216.81,108.41);
\definecolor{drawColor}{gray}{0.20}

\path[draw=drawColor,line width= 0.6pt,line join=round] ( 32.77, 24.58) --
	( 32.77, 27.33);

\path[draw=drawColor,line width= 0.6pt,line join=round] ( 51.66, 24.58) --
	( 51.66, 27.33);

\path[draw=drawColor,line width= 0.6pt,line join=round] ( 70.55, 24.58) --
	( 70.55, 27.33);

\path[draw=drawColor,line width= 0.6pt,line join=round] ( 89.45, 24.58) --
	( 89.45, 27.33);

\path[draw=drawColor,line width= 0.6pt,line join=round] (108.34, 24.58) --
	(108.34, 27.33);

\path[draw=drawColor,line width= 0.6pt,line join=round] (127.23, 24.58) --
	(127.23, 27.33);

\path[draw=drawColor,line width= 0.6pt,line join=round] (146.13, 24.58) --
	(146.13, 27.33);

\path[draw=drawColor,line width= 0.6pt,line join=round] (165.02, 24.58) --
	(165.02, 27.33);

\path[draw=drawColor,line width= 0.6pt,line join=round] (183.91, 24.58) --
	(183.91, 27.33);

\path[draw=drawColor,line width= 0.6pt,line join=round] (202.81, 24.58) --
	(202.81, 27.33);
\end{scope}
\begin{scope}
\path[clip] (  0.00,  0.00) rectangle (216.81,108.41);
\definecolor{drawColor}{RGB}{0,0,0}

\node[text=drawColor,anchor=base,inner sep=0pt, outer sep=0pt, scale=  0.80] at ( 32.77, 16.87) {1};

\node[text=drawColor,anchor=base,inner sep=0pt, outer sep=0pt, scale=  0.80] at ( 51.66, 16.87) {2};

\node[text=drawColor,anchor=base,inner sep=0pt, outer sep=0pt, scale=  0.80] at ( 70.55, 16.87) {3};

\node[text=drawColor,anchor=base,inner sep=0pt, outer sep=0pt, scale=  0.80] at ( 89.45, 16.87) {4};

\node[text=drawColor,anchor=base,inner sep=0pt, outer sep=0pt, scale=  0.80] at (108.34, 16.87) {5};

\node[text=drawColor,anchor=base,inner sep=0pt, outer sep=0pt, scale=  0.80] at (127.23, 16.87) {6};

\node[text=drawColor,anchor=base,inner sep=0pt, outer sep=0pt, scale=  0.80] at (146.13, 16.87) {7};

\node[text=drawColor,anchor=base,inner sep=0pt, outer sep=0pt, scale=  0.80] at (165.02, 16.87) {8};

\node[text=drawColor,anchor=base,inner sep=0pt, outer sep=0pt, scale=  0.80] at (183.91, 16.87) {9};

\node[text=drawColor,anchor=base,inner sep=0pt, outer sep=0pt, scale=  0.80] at (202.81, 16.87) {10};
\end{scope}
\begin{scope}
\path[clip] (  0.00,  0.00) rectangle (216.81,108.41);
\definecolor{drawColor}{RGB}{0,0,0}

\node[text=drawColor,anchor=base,inner sep=0pt, outer sep=0pt, scale=  0.80] at (117.79,  7.06) {Number of Clusters};
\end{scope}
\begin{scope}
\path[clip] (  0.00,  0.00) rectangle (216.81,108.41);
\definecolor{drawColor}{RGB}{0,0,0}

\node[text=drawColor,rotate= 90.00,anchor=base,inner sep=0pt, outer sep=0pt, scale=  0.80] at ( 11.01, 65.12) {Total Within-Cluster Sum of Squares (WSS)};
\end{scope}
\end{tikzpicture}

    \caption{
        Scree plot for clustering analysis of questionnaire variables. The Within-Cluster Sum of Squares (WCSS) score is shown on the y-axis, and the number of clusters is shown on the x-axis.
    }
    \label{fig:scree_plot}

\end{figure}
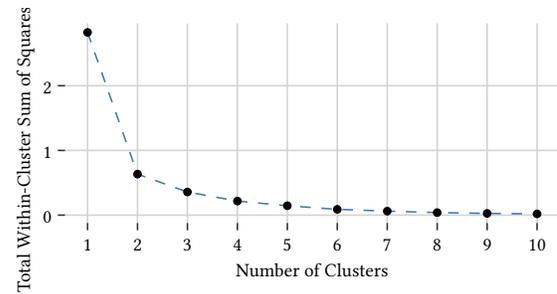

K-means clustering was applied to variables in the two-factor space, with the optimal number of clusters determined via the elbow method and silhouette analysis. The elbow method (Figure \ref{fig:scree_plot}) suggested two to three clusters, while silhouette analysis peaked at 0.65 with two clusters. A three-cluster solution had a lower silhouette score (0.54) and resulted in a single-variable cluster containing Question \ref{q:toomuchinfo}, making it less interpretable. A two-cluster solution was retained.

\section{Apathy toward ethical considerations}\label{app:apathy}
A small percentage of participants (14 out of 303 comments on \ref{q:influence}, 4.62\%) expressed clear apathy regarding ethical aspects of their purchasing decisions. Despite potential social pressure, these participants openly shared their disinterest or skepticism about the relevance of ethical concerns when choosing products, focusing instead on factors like price or brand reputation.

\begin{center}
    \begin{tabular}{p{.9\linewidth}}
        \midrule
        \post ``\textit{Not at all. I rarely ever consider child labour or employment rights when I buy things. Most products are manufactured in China so it is pretty much a given that forced labour has at one point happend.}'' (Participant 189, Low EMCB) \\
    \end{tabular}
    \begin{tabular}{p{.9\linewidth}}
        \post ``\textit{Not that much as I'm not into ideologies.}'' (Participant 76, High EMCB) \\
    \end{tabular}
    \begin{tabular}{p{.9\linewidth}}
        \post ``\textit{I am only interested in the brand name and product quality. DEI policies have no influence on my decision making.}'' (Participant 88, High EMCB) \\
        \midrule
        \multicolumn{1}{r}{Examples of Apathy}
    \end{tabular}
\end{center}

These comments show that for some, ethical concerns are secondary, often overshadowed by more practical considerations.

\end{document}